\documentclass[a4paper,11pt]{article}
\usepackage{jheppub}
\makeatletter
\gdef\@fpheader{}
\usepackage{jheppub}
\usepackage[T1]{fontenc}
\usepackage{graphicx}
\usepackage{amsmath,amssymb,amsfonts}
\usepackage{xspace}
\usepackage{subfig}
\usepackage{float}
\usepackage{color}
\usepackage{xcolor}
\usepackage{bbding}
\usepackage[utf8]{inputenc}
\usepackage{commath}
\usepackage{slashed}
\usepackage{fancyvrb}
\usepackage[utf8]{inputenc}
\usepackage[english]{babel}
\usepackage{orcidlink}
\usepackage{cancel}
\usepackage{hyperref}
\usepackage{tabularray}
\UseTblrLibrary{booktabs}
\usepackage{gensymb}
\usepackage{scalerel}
\usepackage[titletoc]{appendix}
\usepackage{comment}
\usepackage{enumitem}

\title{Singlet-Doublet fermion origin of dark matter, neutrino mass and inverse first-order electroweak phase transition}

\author[a]{Debasish Borah$^{\orcidlink{https://orcid.org/0000-0001-8375-282X}}$}
\emailAdd{dborah@iitg.ac.in}

\author[a]{, Indrajit Saha$^{\orcidlink{https://orcid.org/0000-0002-7459-0838}}$}
\emailAdd{s.indrajit@iitg.ac.in}

\affiliation[a]{Department of Physics, Indian Institute of Technology Guwahati, Assam 781039, India}
\author[b]{, Sujit Kumar Sahoo$^{\orcidlink{https://orcid.org/0000-0002-9014-933X}}$}
\emailAdd{ph21resch11008@iith.ac.in}

\author[b]{, Narendra Sahu$^{\orcidlink{https://orcid.org/0000-0002-9675-0484}}$}
\emailAdd{nsahu@phy.iith.ac.in}

\author[b]{, Shashwat Sharma$^{\orcidlink{https://orcid.org/0000-0002-9675-0484}}$}
\emailAdd{ph23resch11016@iith.ac.in}
\affiliation[b]{Department of Physics, Indian Institute of Technology Hyderabad, Kandi, Telangana-502285, India.}
\abstract{We study the possibility of an inverse first-order electroweak phase transition (IFOEWPT) and observable gravitational waves (GW) in a radiative neutrino mass model of scotogenic type where singlet-doublet (SD) fermions, the lightest of whom is the dark matter (DM) candidate, generate the necessary seesaw at one-loop level. Considering the possibility of light neutrinos being Dirac for simplicity and additional detection prospects, we extend the standard model (SM) with two generations of $SU(2)_L$ singlet and doublet fermions, one singlet scalar, and three right-handed neutrinos (RHNs). While RHNs provide the right chiral parts of light Dirac neutrinos, the SD fermions and the scalar singlet facilitate the one-loop neutrino mass diagram. The neutral component of the lighter SD fermion, stabilized under a residual $Z_2$ symmetry plays the role of DM while the heavier SD fermions strongly couple to the Higgs leading to an IFOEWPT where the Universe undergoes two different first-order phase transition as it goes from the symmetric to the final broken Higgs phase. We constrain the parameter space from the requirements of generating the correct neutrino mass, DM relic as well as IFOEWPT while incorporating the existing constraints from different experiments. The final allowed parameter space of the model can be probed at collider, direct-detection, GW and cosmic microwave background (CMB) experiments in near future.}
\begin{document}
\maketitle
\flushbottom

\section{Introduction}
\label{sec:intro}
The matter component in the present Universe is dominated by a non-luminous, non-baryonic form of matter, popularly referred to as the dark matter (DM) \cite{Planck:2018vyg,ParticleDataGroup:2024cfk,Cirelli:2024ssz}. Assuming DM to have particle origin, none of the standard model (SM) particles satisfy the required criteria for a particle DM. This has motivated several beyond standard model (BSM) studies out of which the weakly interacting massive particle (WIMP) \cite{Kolb:1990vq, Jungman:1995df, Bertone:2004pz} remains the most popular one. In the WIMP framework, DM can be thermally produced in the early Universe due to its sufficiently large non-gravitational interactions with the SM bath. Subsequently, DM relic is set by its thermal freeze-out. The same DM-SM interactions can also lead to observable DM-nucleon scattering at terrestrial detectors like \texttt{LUX-ZEPLIN (LZ)} ~\cite{LZ:2024zvo}, \texttt{XENONnT} \cite{XENON:2025vwd}, \texttt{PandaX-4T} \cite{PandaX:2024qfu} keeping the WIMP scenario predictive and verifiable. Another observed phenomena which the SM fails to address is the light neutrino mass and mixing, which has been confirmed by neutrino oscillation experiments \cite{ParticleDataGroup:2024cfk}. Non-zero neutrino mass and mixing can be accommodated within seesaw models like seesaw models. While canonical high scale seesaw models like type-I \cite{Minkowski:1977sc,Yanagida:1979as,Gell-Mann:1979ijt, GellMann:1980vs,Mohapatra:1979ia,Schechter:1980gr,Schechter:1981cv}, type-II \cite{Mohapatra:1980yp, Schechter:1981cv, Wetterich:1981bx, Lazarides:1980nt, Brahmachari:1997cq} and type-III \cite{Foot:1988aq} seesaw have been widely studied in the literature, radiative seesaw models\footnote{See \cite{Cai:2017jrq} for a review of radiative seesaw models.} like the scotogenic model \cite{Tao:1996vb, Ma:2006km} have received significant attention recently as it offers a common origin of DM and light neutrino mass.

A particularly economical and predictive WIMP scenario is the singlet–doublet dark matter (SDDM) model \cite{Bhattacharya:2018fus,Cynolter:2015sua,Bhattacharya:2015qpa,Bhattacharya:2017sml,Bhattacharya:2018cgx,Bhattacharya:2016rqj,Dutta:2020xwn,Borah:2021khc,Borah:2021rbx,Borah:2022zim,Borah:2023dhk,Paul:2024iie,Paul:2026snc,DEramo:2007anh,Cohen:2011ec,Freitas:2015hsa,Calibbi:2015nha,Cheung:2013dua,Banerjee:2016hsk,DuttaBanik:2018emv,Horiuchi:2016tqw,Restrepo:2015ura,Abe:2017glm,Konar:2020wvl,Konar:2020vuu,Calibbi:2018fqf,Ghosh:2021wrk,Kundu:2024nig,Bhattacharya:2021ltd,Enberg:2007rp,Oncala:2021tkz,Paul:2024prs,Paul:2025spm,Dey:2025pcs}, in which the dark sector contains a fermionic singlet $N$ and an electroweak vector-like doublet $\Psi$. The DM emerges as an admixture of the neutral component of $\Psi$ and $N$ after the electroweak symmetry breaking (EWSB). This setup is characterized by just three main parameters: the dark matter mass, the singlet–doublet mixing angle, and the mass splitting between the DM and heavier dark sector state. The model yields viable dark matter candidates across a wide mass range from the GeV to the TeV scale. The minimal SDDM framework can be augmented by adding a singlet scalar $\phi$, allowing Majorana neutrino masses to be generated radiatively at one loop \cite{Fraser:2014yha,Konar:2020wvl}. Similar to the original scotogenic model, the particles inside the loop namely the SDDM and $\phi$ transform non-trivially under an unbroken $Z_2$ symmetry such that the lightest $Z_2$-odd particle can be DM. In order to generate solar and atmospheric neutrino mass scales, one needs at least two copies of the singlet scalar $\phi$ or two copies of singlet-doublet $N-\Psi$. It is also possible to have light Dirac neutrino mass generated radiatively from SDDM by appropriately incorporating right chiral parts of neutrinos and preventing lepton number violating Majorana mass terms with appropriate symmetries \cite{Borah:2023dhk}. Dirac nature of light neutrinos bring additional detection aspects like enhanced dark radiation $\Delta N_{\rm eff}$ and absence of neutrinoless double beta decay signatures.

Motivated by this, here we consider an interesting signature of SDDM model of radiative Dirac neutrino mass by incorporating the role of the singlet-doublet (SD) fermions on electroweak symmetry breaking in the early Universe. While a first-order EWPT is not possible in the SM alone, several BSM extensions can naturally accommodate it. Conventionally, most of these extensions have focused on scalar extensions of the SM due to the natural occurance of a barrier in the Higgs effective potential at high temperatures \cite{Dey:2025pcs}. However, a few of these BSM proposals also considered the role of fermions alone in ensuring a first-order EWPT \cite{Carena:2004ha, Davoudiasl:2012tu, Egana-Ugrinovic:2017jib, Baldes:2016gaf, Baldes:2016rqn, Braconi:2018gxo, Angelescu:2018dkk, Cao:2021yau, Ai:2025vfi}. These works considered new fermions heavier than the critical temperature and having large couplings to the SM Higgs. In particular, the authors of \cite{Egana-Ugrinovic:2017jib, Angelescu:2018dkk, Ai:2025vfi} considered singlet-doublet fermions with large couplings to the SM Higgs. While usual first-order EWPT involves tunneling from the symmetric to the broken phase, the one driven by SD fermion follows an unusual trajectory in the field space, referred to as the inverse first-order EWPT (IFOEWPT). As the Universe cools, the symmetric phase at high temperature goes through a crossover first, followed by two different FOPTs in order for the SM Higgs to settle to its true minimum. We consider two generations of vector-like singlet-doublet fermions required for generating correct neutrino masses. The additional singlet scalar present for radiative neutrino mass does not play any role in FOPT due to the chosen Higgs portal couplings. One copy of the SD fermions plays dominant role in FOPT while the other plays the role of WIMP DM with both of them playing non-trivial roles in generating light neutrino masses. While this can be utilized for radiative Majorana neutrino mass scenario too, we consider a Dirac radiative seesaw \cite{Borah:2023dhk} in this work. We find the parameter space consistent with all phenomenological constraints related to DM, neutrino mass, colliders, $\Delta N_\text{eff}$ and project the currently allowed region to be reached by future gravitational wave (GW) experiments.

This paper is organized as follows. In section \ref{sec:model}, we discuss our model in details followed by dark matter analysis in section \ref{sec:DiracDM}. In section \ref{sec:InvPT}, we discuss the inverse first-order EWPT driven by SD fermions. We briefly discuss the possibility of enhanced $\Delta N_{\rm eff}$ and collider signatures of singlet-doublet fermions in section \ref{sec:Neff} and section \ref{sec:Collider} respectively. Finally, we summarize our results and conclude in section \ref{sec:results}.


\section{Radiative Dirac Seesaw with Singlet-Doublet Dark Matter}
\label{sec:model}
We extend the SM by introducing two generations of vector-like singlet fermion $\chi_i$,  doublet fermion $\Psi_i$, one complex scalar $\phi(=\phi_1+i\phi_2)$, and three generations of Dirac right-handed neutrinos $\nu_R$. A discrete $\mathcal{Z}_4$ symmetry is imposed under which the fermions $\chi_i$ and $\psi_i$ carry charge $-1$, while the scalar fields $\phi_i$ carry charge $i$. The SM lepton doublets $L_L$ and charged leptons $\ell_R$ transform with charge $-i$, whereas the Dirac right-handed neutrinos $\nu_R$ have charge $i$. The respective charge assignments are given in Table~\ref{tab:tab1}. Note that at least two generations of SD fermions are required to generate two non-zero neutrino mass eigenstates to satisfy neutrino oscillation data.

\begin{table}[h!]
		\small
		\begin{center}
			\begin{tabular}{||@{\hspace{0cm}}c@{\hspace{0cm}}|@{\hspace{0cm}}c@{\hspace{0cm}}|@{\hspace{0cm}}c@{\hspace{0cm}}|@{\hspace{0cm}}c@{\hspace{0cm}}||}
				\hline
				\hline
				\begin{tabular}{c}
                {\bf ~~~~Symmetry~~~~}\\
					{\bf ~~~~Group~~~~}\\ 
					\hline
					
					$SU(2)_{L}$\\ 
					\hline
					$U(1)_{Y}$\\ 
					\hline
					$\mathcal{Z}_4$\\ 
				\end{tabular}
				&
				&
				\begin{tabular}{c|c|c|c|c}
					\multicolumn{5}{c}{\bf Fermion Fields}\\
					\hline
					~~~$L_L$~~~&~~~$\ell_R$~~~& ~~~$\Psi_i =(\psi_i^0~~\psi_i^-)^T $~~~ & ~~~$\chi_i$~~~&~~~$\nu_R$~~~ \\
					\hline
					$2$&$1$&$2$&$1$&$1$\\
					\hline
					$-1$&$-2$&$-1$&$0$&$0$\\
					\hline
					$-i$&$-i$&$-1$&$-1$&$i$\\
				\end{tabular}
				&
				\begin{tabular}{c|c|c}
					\multicolumn{2}{c}{\bf Scalar Field}\\
					\hline
					~~~$H$~~~& ~~~$\phi$~~~\\
					\hline
					$2$&$1$\\
					\hline
					$1$&$0$\\
                    \hline
					$+1$&$i$\\
				\end{tabular}\\
				\hline
				\hline
			\end{tabular}
			\caption{}
			\label{tab:tab1}
		\end{center}    
	\end{table}

\noindent
The relevant Lagrangian terms and the interactions among the particles are given as
\begin{eqnarray}
\label{eq:lag}
    \mathcal{L}\supset && i\bar{\Psi_i}\gamma^{\mu}D_{\mu}\Psi_i +i\bar{\chi_i}\gamma^{\mu}D_{\mu}\chi_i + (D_{\mu}\phi)^*(D^{\mu}\phi)  - M_{\Psi_{i}}\bar{\Psi_i}\Psi_i - M_{\chi_{i}} \bar{\chi_i}\chi_i \nonumber  \\  
    &&- \lambda_{\Psi_{i \alpha}}\bar{L}_{L_{\alpha}}\phi \Psi_i  - y_{i}\bar{\Psi}_i\tilde{H}\chi_i - \lambda_{\chi_{i \beta}} \bar{\chi_i} \phi \nu_{R_{\beta}} + h.c.-V(\phi,H),
\end{eqnarray}
where the most general scalar potential is given as
\begin{eqnarray}
    V(\phi,H)=&&-\mu_{H}^2(H^\dagger H)+\lambda_H (H^\dagger H)^2 +\lambda_{\phi H}(\phi^\dagger\phi)(H^\dagger H) \nonumber\\
    &&+ M_{\phi}^2(\phi^\dagger\phi) +\lambda_{\phi}(\phi^\dagger\phi)^2.
\end{eqnarray}
In order to generate the Dirac mass for neutrinos, the $\mathcal{Z}_4$ symmetry must be broken. We achieve this by introducing a soft symmetry-breaking term in the scalar potential $\frac{1}{2}\mu_{\phi}^2 \left(\phi^2 + (\phi^\dagger)^2 \right),$ which explicitly breaks the $\mathcal{Z}_4$ symmetry. After EWSB, the SM Higgs acquires a vacuum expectation value (VEV). 
\begin{eqnarray}
		H=\begin{pmatrix}
			0\\
			\frac{v+h}{\sqrt{2}}
		\end{pmatrix}.
	\end{eqnarray}
In the presence of this soft $\mathcal{Z}_4$-breaking term, the complex scalar singlet $\phi$ splits into two physical scalars having masses given by
\begin{eqnarray}
    M_{\phi_{1}}^2 &=& M_{\phi}^2 + \mu_{\phi}^2 + \frac{1}{2}\lambda_{\phi H} v^2, \\
    M_{\phi_{2}}^2 &=& M_{\phi}^2 - \mu_{\phi}^2 + \frac{1}{2}\lambda_{\phi H} v^2.
\end{eqnarray}
The corresponding mass-squared splitting is $
\Delta M_{\phi}^2 = M_{\phi_{1}}^2 - M_{\phi_{2}}^2 = 2\mu_{\phi}^2$. We will demonstrate in section.~\ref{sec:diracnumass} that this mass splitting plays a crucial role in generating Dirac neutrino masses at the one-loop level. Furthermore, the Higgs VEV ($v$) induces mixing between the singlet fermion and the neutral component of the doublet fermion through the interaction $\bar{\Psi_i}\tilde{H}\chi_i$, leading to singlet--doublet Dirac dark matter\footnote{It should be noted that the $\mathcal{Z}_4$ symmetry alone can not ensure the Dirac nature of DM and hence neutrinos. We assume the presence of an unbroken global $U(1)$ symmetry, similar to lepton number which keeps all the Majorana mass terms away.}. Due to the mixing between $\chi_i$ and $\psi_i^0$, parametrized by $\sin\theta_i$, we get two neutral mass eigenstates. We denote the first generation mass eigenstates as: $\chi$ and $\psi$ with the mixing angle $\sin\theta_1$, and the second generation mass eigenstates as: $\chi'$ and $\psi'$ with the mixing angle $\sin\theta_2$. Consequently, the transformation from the flavor states to physical states can be written as,
\begin{align}
    \begin{pmatrix}
        \chi_1 \\
        \psi^0_1
    \end{pmatrix}=
    \begin{pmatrix}
			\cos\theta_1 & \sin\theta_1 \\
			-\sin\theta_1 & \cos\theta_1
		\end{pmatrix}
        \begin{pmatrix}
        \chi \\
        \psi
    \end{pmatrix}, ~~~\text{and}~~~
    \begin{pmatrix}
        \chi_2 \\
        \psi^0_2
    \end{pmatrix}=
    \begin{pmatrix}
			\cos\theta_2 & \sin\theta_2 \\
			-\sin\theta_2 & \cos\theta_2
		\end{pmatrix}
        \begin{pmatrix}
        \chi' \\
        \psi'
    \end{pmatrix}.
\end{align}
A relationship between the mixing angle $\theta_i$ and the Yukawa coupling $y_i$ can be established as:
\begin{align}
    y_i=\frac{(M_{\Psi_i}-M_{\chi_i})\tan2\theta_i}{\sqrt{2}v}.
\end{align}
The mass eigenvalues of the physical states are given as
\begin{align}
    M_{X}&=\frac{1}{2}\left(M_{\chi_i}+M_{\Psi_i}-\sqrt{(M_{\chi_i}-M_{\Psi_i})^2+4v^2y_i^2}\right)\\
    M_{Y}&=\frac{1}{2}\left(M_{\chi_i}+M_{\Psi_i}+\sqrt{(M_{\chi_i}-M_{\Psi_i})^2+4v^2y_i^2}\right),
\end{align}
where $X\in\{\chi,\chi'\}$, $Y\in\{\psi,\psi'\}$ represents the fermion mass eigenstates, and $i$ represents fermion generation index (e.g. for $i=1$, $\{X,Y\}=\{\chi,\psi\}$, similarly, for $i=2$, $\{X,Y\}=\{\chi',\psi'\}$). 
From Eq.~\eqref{eq:lag}, the interaction among the mass eigenstates ($\chi$, $\psi^\pm_1$ and $\psi$), can be expressed as,
\begin{eqnarray}
    \label{eq:intlagragian}
    \mathcal{L}_{\rm int}&\supset&\frac{-e_o}{2\sin\theta_W\cos\theta_W}\left[\sin^2\theta_i \bar{X}\gamma^\mu Z_\mu X + \cos^2\theta_i \bar{Y}\gamma^\mu Z_\mu Y -\frac{1}{2}\sin2\theta_i \left(\bar{X}\gamma^\mu Z_\mu Y +\bar{Y}\gamma^\mu Z_\mu X\right) \right]\nonumber \\
    &+&\frac{e_o}{\sqrt{2}\sin\theta_W}\left[ \sin\theta_i\left( \bar{\psi}^-_i\gamma^\mu W_\mu^-X +  \bar{X}\gamma^\mu W_\mu^+\psi^-_i \right) -\cos\theta_i\left( \bar{\psi}^-_i\gamma^\mu W_\mu^-Y +  \bar{Y}\gamma^\mu W_\mu^+\psi_i^- \right) \right]\nonumber\\
    &+&\frac{e_o}{2\sin\theta_W\cos\theta_W}\cos2\theta_i\, \bar{\psi}^-_i\gamma^\mu Z_\mu\psi_i^- +e_o\bar{\psi}^-_i\gamma^\mu A_\mu\psi_i^- +\frac{y_ih}{\sqrt{2}}\left[ \sin2\theta_i\left( \bar{X}X-\bar{Y}Y \right)\right.\\
    &-&\left.\cos2\theta_i \left( \bar{X}Y+\bar{Y}X \right) \right]+\sin\theta_i \left[ \left( \lambda_{\psi_{i\alpha}} \bar{\nu}_{L_\alpha}\phi X +\lambda^*_{\psi_{i\alpha}} \bar{X}\phi^* \nu_{L_\alpha}  \right) -\left( \lambda_{\chi_{i\beta}} \bar{Y}\phi \nu_{R_\beta} +\lambda^*_{\chi_{i\beta}} \bar{\nu}_{R_\beta}\phi^* Y   \right) \right]\nonumber\\
    &-&\cos\theta_i \left[ \left( \lambda_{\psi_{i\alpha}} \bar{\nu}_{L_\alpha}\phi Y +\lambda^*_{\psi_{i\alpha}} \bar{Y}\phi^* \nu_{L_\alpha}  \right) +\left( \lambda_{\chi_{i\beta}} \bar{X}\phi \nu_{R_\beta} +\lambda^*_{\chi_{i\beta}} \bar{\nu}_{R_\beta}\phi^* X   \right) \right]\nonumber\\
    &-&\lambda_{\psi_{i\alpha}}\bar{\ell}_\alpha^-\phi\psi_i^- - \lambda^*_{\psi_{i\alpha}}\bar{\psi}^-_i\phi\ell_\alpha^-\nonumber
\end{eqnarray}
    
where $e_0=0.313$ is the electromagnetic coupling constant, and $\theta_W$ is the Weinberg angle. In this setup, we consider a mass hierarchy: $M_{\chi}<M_\psi,M_{\psi^\pm},M_{\phi_{1,2}}, M_{\chi'},M_{\psi'},M_{\psi'^\pm}$ and $M_{\phi_{1,2}}<M_{\chi'},M_{\psi'},M_{\psi'^\pm}$. $\chi$ being the lightest among the dark sector particles serves as DM candidate. The first condition ensures the stability of DM, while the latter ensures the decay of second generation fermions to the lighter dark sector particles.

\subsection{Radiative Neutrino Mass}
\label{sec:diracnumass}
\begin{figure}
    \centering
    \includegraphics[width=0.6\linewidth]{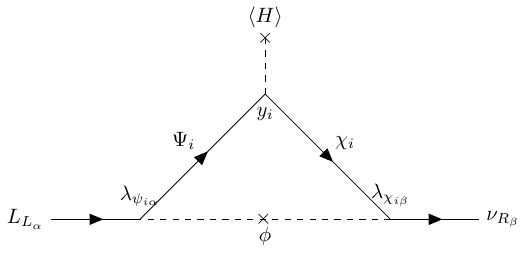}
    \caption{Dirac Neutrino mass at one loop}
    \label{fig:neutrinomassDirac}
\end{figure}

As discussed in the previous section, the $\mathcal{Z}_4$ symmetry is explicitly broken by the soft term $ \frac{1}{2}\mu_{\phi}^2 \left(\phi^2 + (\phi^\dagger)^2 \right).
$ As a result, the Dirac neutrino mass operator $\bar{L}\tilde{H}\nu_R$ is generated at the one-loop level. This arises from loop diagrams involving the two generations of SD fermions $(\psi_i^0, \chi_i)$ and the singlet scalar field $\phi$, as illustrated in Fig.~\ref{fig:neutrinomassDirac}.
The one-loop neutrino mass can be calculated as:
\begin{eqnarray}
    \label{eq:neutrinomass}
    m_{\nu_{\alpha \beta}} =\frac{-i \mu_{\phi}^2}{4 \pi^2}\sum_i (\lambda_{\Psi_{i\alpha}})^T\left( ((M_Y - M_X) \sin{2\theta_i}) F(M_{X},M_{Y},M_{\phi_1},M_{\phi_2})_{i}I_{ii} \right)(\lambda_{\chi_{i\beta}})
\end{eqnarray}
where  $X\in\{\chi,\chi'\}$, $Y\in\{\psi,\psi'\}$ represents the fermion mass eigenstates, and $i$ represents the generation index of the SD, $\alpha$ and $\beta$ represent lepton flavour indices, $I_{ii}$ is the $2\times2$ identity matrix and $F(M_{X},M_{Y},M_{\phi_1},M_{\phi_2})_{i} $ is the loop factor and is given by
\begin{eqnarray}
\label{eq:loopfactor}    F(M_{X},M_{Y},M_{\phi_1},M_{\phi_2})_i=&&\frac{(M_{X}+M_{Y})M_{Y}^3 \ln{\left( \frac{M_{Y}^2}{M_{X}^2} \right)}}{(M_{X}^2-M_{Y}^2)(M_{\phi_2}^2-M_{Y}^2)(M_{\phi_1}^2-M_{Y}^2)} +  \frac{1}{(M_{\phi_{1}}^2-M_{\phi_{2}}^2)} \\&& \times \left( \frac{M_{\phi_{2}}^2(M_{\phi_{2}}^2+M_{X}M_{Y})\ln{\left( \frac{M_{\phi_{2}}^2}{M_{X}^2} \right)}}{(M_{\phi_{2}}^2-M_{X}^2)(M_{\phi_{2}}^2-M_{Y}^2)}  - \frac{M_{\phi_{1}}^2(M_{\phi_{1}}^2+M_{X}M_{Y})\ln{\left( \frac{M_{\phi_{1}}^2}{M_{X}^2} \right)}}{(M_{\phi_{1}}^2-M_{X}^2)(M_{\phi_{1}}^2-M_{Y}^2)} \right).\nonumber
\end{eqnarray}
Since the Dirac neutrino mass is defined as Eq.~\eqref{eq:neutrinomass}, we can define a biunitary transformation to diagonalize the neutrino mass matrix as
\begin{eqnarray}
    \label{eq:diagnumassdirac}
    D_{m_{3\times3}} = (V_{L_{3\times3}})^\dagger m_{\nu_{3\times 3}} V_{R_{3\times3}}.
\end{eqnarray}
Now we can further write the Eq.~\eqref{eq:diagnumassdirac} as
\begin{eqnarray}
    \label{eq:diracparameterization}
    D_{m_{ii}}=(V_{L_{\alpha i}})^\dagger(\lambda_{\psi_{k\alpha}})^T\Lambda_{kk} \lambda_{\chi_{k\beta}}V_{R_{\beta i}},
\end{eqnarray}
where $\Lambda$ is a $2\times2$ diagonal matrix with its diagonal elements represented by
\begin{eqnarray}
    \Lambda_{kk} = \frac{ \mu_{\phi}^2}{4 \pi^2}\left( (\Delta M_k \sin{2\theta_k}) F(M_{X},M_{Y},M_{\phi_{1}},M_{\phi_{2}}) \right).
\end{eqnarray}
Now using Eq.~\eqref{eq:diracparameterization}, we can parameterize the couplings $\lambda_{\chi_{i\alpha}}$ and $\lambda_{\psi_{i\alpha}}$ as below
\begin{eqnarray}           
(\lambda_{\psi})^T&=& V_L~D_{\sqrt{m}}^T~R^{-1}~P^T~\sqrt{\Lambda^{-1}}, \nonumber\\
\lambda_{\chi}&=&\sqrt{\Lambda^{-1}}~P~R~ D_{\sqrt{m}}~V_R^{\dagger}\label{eq:ci}
\end{eqnarray}
where $R$ is a general $3\times3$ complex matrix, and $P$ is a projection matrix defined as:
\begin{eqnarray}
    \label{eq:projection}
    P=
    \begin{pmatrix}
        0&1&0\\
        0&0&1
    \end{pmatrix}.
\end{eqnarray}
$V_L$ can be treated as the left-handed neutrino mixing matrix or the Pontecorvo--Maki--Nakagawa--Sakata (PMNS) leptonic mixing matrix $U_{\rm PMNS}$ and $V_R$ is some general unitary matrix (which we consider as identity here). Thus, the Yukawa coupling matrices $\lambda_\psi$ and $\lambda_{\chi}$ are $2\times3$ matrices, where the index 2 represents the SD fermion generation index and the index 3 represents the lepton flavor index. For simplicity, with the chosen form of the projection matrix $P$ and $V_R$, the electron-type coupling of $\lambda_\chi$, given in Eq.~\eqref{eq:ci}, is set to zero.


\subsection{Muon anomalous magnetic moment}
\label{sbsec:muong-2Dirac}
\begin{figure}
    \centering
    \includegraphics[scale=1.3]{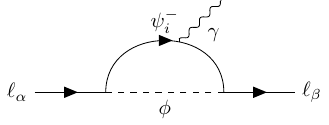}
    \caption{The Feynmann diagram giving rise $(g-2)$ and charged lepton flavor violation.}
    \label{fig:lfvDirac}
\end{figure}

In our setup, the new positive contribution to the muon $(g-2)$ arises from the one-loop diagram involving the charged component of the doublet fermion $\psi^{-}_i$ and the singlet scalar $\phi$ running in the loop, as shown in Fig.~\ref{fig:lfvDirac}. The corresponding estimate is given by
\begin{eqnarray}
    \label{eq:g-2Dirac}
    \Delta a_{\mu}=\frac{m_{\mu}^2}{(4\pi)^2}\sum_{i=1}^3(\lambda_{\psi_{i\mu}}^*\lambda_{\psi_{i\mu}})\int_{0}^1 dx \frac{(1-x)^2(x+\frac{m_{\psi_i^-}}{m_\mu})}{(1-x)(m_{\psi_i^-}^2 -x\, m_{\mu}^2)+x\,m_{\phi}^2}.
\end{eqnarray}
The present value of the muon anomalous magnetic moment is reported to be $\Delta{a_{\mu}}=38(63)\times10^{-11}$ \cite{Muong-2:2025xyk,Aliberti:2025beg}. This indicates that, within the current uncertainties, there is no statistically significant tension between the experimental measurement and the SM prediction. Nevertheless, we will use the upper limit on $\Delta{a_{\mu}}=101\times10^{-11}$ to constrain the model parameter space in the upcoming sections. 

\subsection{Charged lepton flavor violation}
\label{sbsec:lfvDirac}
\begin{figure}
    \centering
    \includegraphics[scale=0.45]{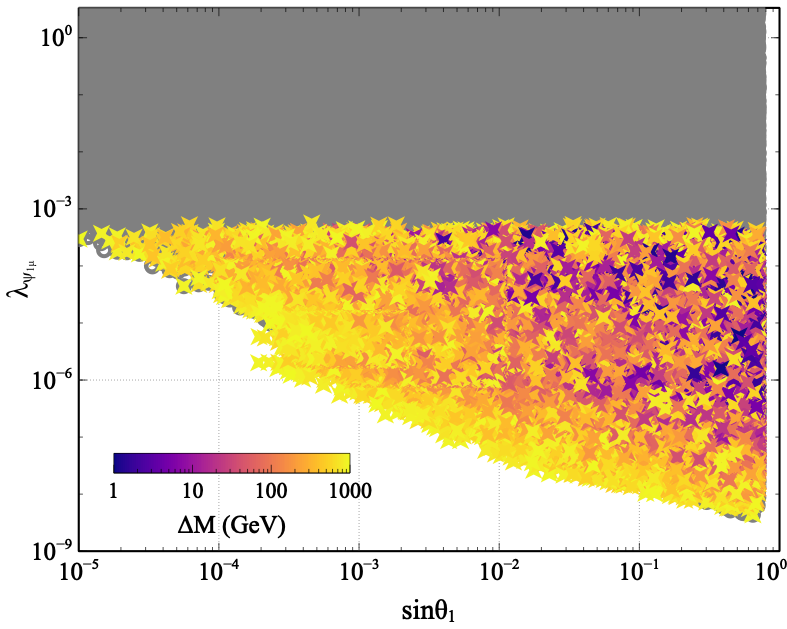}
    \caption{ The parameter space allowed by $(g-2)_\mu$, cLFV and neutrino oscillation data is shown by the colored points in the plane of $\sin\theta_1$ and $\lambda_{\psi_{1\mu}}$, while the gray colored points satisfy only the neutrino oscillation data.}
    \label{fig:lfvparams}
\end{figure}

Charged lepton flavor–violating (cLFV) processes can occur at the one-loop level, although they are extremely suppressed by the tiny neutrino masses, in the absence of any new degrees of freedom. In our framework, the presence of the singlet scalar and the doublet fermions leads to additional cLFV contributions through the loop-mediated diagrams shown in Fig.~\ref{fig:lfvDirac}. The branching ratio for the process $\alpha\rightarrow \beta\gamma$ is given by \cite{Lindner:2016bgg},
\begin{eqnarray}
    \label{eq:LFVDirac}
    {\rm Br}(\alpha \rightarrow \beta\gamma)&\approx & \frac{3(4\pi)^3 \alpha_{em}}{4 G_F^2} \\ 
    &\times& \left\lvert \sum_{i=1}^3  \frac{\lambda_{\psi_{i\alpha}}\lambda_{\psi_{i\beta}}^*}{(4\pi)^2} \int_0^1 dx \int_{0}^{1-x}dy \frac{x(y+(1-x-y)\frac{m_\beta}{m_\alpha})+(1-x)\frac{m_{\psi_i^-}}{m_\alpha}}{-x\,y\,m_{\alpha}^2 -x(1-x-y)m_{\beta}^2+x m_{\phi}^2 +(1-x)m_{\psi_i^-}^2} \right\rvert^2 \nonumber
\end{eqnarray}
where $\alpha,\beta\in\{e,\mu,\tau\}$ represents lepton flavor index.
In Fig.~\ref{fig:lfvparams}, we show the parameter space consistent with neutrino mass generation, the muon anomalous magnetic moment $(g-2)\mu$, and cLFV constraints considering the processes $\mu\rightarrow e\gamma$, $\tau\rightarrow e\gamma$ and $\tau\rightarrow \mu\gamma$, is shown by colored points in the $\lambda_{\psi_{1\mu}}$–$\sin\theta_{1}$ plane. The color scale indicates the mass splitting $\Delta M(\equiv M_{\psi}-M_\chi)$. The experimental bounds on the processes considered are reported to be Br$(\mu\rightarrow e\gamma)<1.5\times10^{-13}$ \cite{Oya:2026qex}, Br$(\tau\rightarrow e\gamma)<3.3\times10^{-8}$ \cite{BaBar:2009hkt} and Br$(\tau\rightarrow \mu\gamma)<4.4\times10^{-8}$ \cite{BaBar:2009hkt}. Gray points in the background satisfy the neutrino mass constraint only. We observe that $\lambda_{\psi_{1\mu}}$ decreases with increasing $\sin\theta_1$, reflecting the seesaw structure underlying neutrino mass generation. It is worth noting that the maximum allowed value of $\lambda_{\psi_{1\mu}}$ is $\sim \mathcal{O}(10^{-4})$, while the minimum value of $\sin\theta_1$ permitted by all constraints is $\sim \mathcal{O}(10^{-5})$.  

\section{DM Relic Density}
\label{sec:DiracDM}
As explained before, after the electroweak symmetry breaking, for both generations of singlet-doublet fermions, the neutral component of the doublet mixes with the singlet fermion, giving rise to two Dirac mass eigenstates each ($X\in\{\chi,\chi'\}$ and $Y\in\{\psi,\psi'\}$). $\chi$ being the lightest among the dark sector serves as the DM candidate. The parameters relevant for the relic calculation are the DM mass $M_{\rm DM}$ ($\equiv M_\chi$), the singlet-doublet mass splitting $\Delta{M}$, the singlet-doublet mixing angle $\sin\theta_1$, singlet scalar masses $M_{\phi_1}-M_{\phi_2}$, masses of the heavier singlet-doublet pair and the corresponding mixing angle $\sin\theta_2$, and the Yukawa couplings $\lambda_\psi,\lambda_\chi$. We vary these parameters in the model, keeping a mass hierarchy $M_{\chi}\leq M_{\psi},M_{\phi_{1,2}} \leq M_{\chi'},M_{\psi'}$ to avoid multi-component DM. The Yukawa couplings $\lambda_{\chi}$ and $\lambda_{\psi}$ are calculated using Eq.~\eqref{eq:ci}. 

In order to calculate relic density, we define two dark sectors: (a) sector 1, containing $\chi$, and (b) sector 2, comprising $\psi$, $\psi_1^\pm$ and other dark sector particles ($\chi'$, $\psi'$, $\psi_2^\pm$, and $\phi_{1,2}$), while all SM particles are assigned to sector 0. We define the comoving number densities of sector 1 and sector 2 particles as $Y_1 \equiv n_{\chi}/s$ and $Y_2 \equiv (n_{\psi} + n_{\psi_1^\pm} +n_{\chi'} +n_{\psi'} +n_{\psi_2^\pm} +n_{\phi_{1,2}} )/s$, respectively, and the total DM relic abundance is defined as the sum of sector 1 and sector 2 abundances: $Y_\text{DM}=Y_1+Y_2$. The coupled Boltzmann equations governing their evolution are given as \cite{Paul:2024prs,Alguero:2022inz}
\begin{eqnarray}
		\frac{dY_1}{dT} &=&   \frac{1}{3\mathcal{H}}\frac{ds}{dT} \left[    \langle \sigma_{1100} v \rangle ( Y_1^2 - {Y_1^{\rm eq}}^2) +    \langle \sigma_{1122} v \rangle \left( Y_1^2 - Y_2^2  \frac{{Y_1^{\rm eq}}^2}{{Y_2^{\rm eq}}^2}\right)  + \langle \sigma_{1200} v \rangle ( Y_1 Y_2 - Y_1^{\rm eq}Y_2^{\rm eq})\right. \nonumber\\
		&&+\left.  \langle \sigma_{1222} v \rangle \left( Y_1 Y_2 - Y_2^2   \frac{Y_1^{\rm eq}}{Y_2^{\rm eq}} \right) -\langle \sigma_{1211} v \rangle \left( Y_1 Y_2 - Y_1^2   \frac{Y_2^{\rm eq}}{Y_1^{\rm eq}} \right)
		-\frac{ \Gamma_{2\rightarrow 1}}{s}\left( Y_2 -Y_1 \frac{Y_2^{\rm eq}}{Y_1^{\rm eq}}  \right)        \right], \nonumber 
		\label{eq:Y1}
	\end{eqnarray}
	\begin{eqnarray}
		\frac{dY_2}{dT} &=&   \frac{1}{3\mathcal{H}}\frac{ds}{dT}\left[    \langle \sigma_{2200} v \rangle ( Y_2^2 - {Y_2^{\rm eq}}^2) -    \langle \sigma_{1122} v \rangle \left( Y_1^2 - Y_2^2  \frac{{Y_1^{\rm eq}}^2}{{Y_2^{\rm eq}}^2}\right) +  \langle \sigma_{1200} v \rangle ( Y_1 Y_2 - Y_1^{\rm eq}Y_2^{\rm eq}) \right. \nonumber \\
		&&- \left. \langle \sigma_{1222} v \rangle \left( Y_1 Y_2 - Y_2^2   \frac{Y_1^{\rm eq}}{Y_2^{\rm eq}} \right)
		+\langle \sigma_{1211} v \rangle \left( Y_1 Y_2 - Y_1^2   \frac{Y_2^{\rm eq}}{Y_1^{\rm eq}} \right)  + \frac{ \Gamma_{2\rightarrow 1}}{s}\left( Y_2 -Y_1 \frac{Y_2^{\rm eq}}{Y_1^{\rm eq}}  \right)        \right], \nonumber 
		\label{eq:Y2}
	\end{eqnarray}
    where $Y_i^{\rm eq}\left( \equiv \frac{n_i^{\rm eq}}{s}\right)$ are the equilibrium comiving densities, $\mathcal{H}$, $s=\frac{2\pi^2}{45}g_{*s}T^3$ being the Hubble parameter and entropy density respectively. On the right hand side of the above equations, $\langle \sigma_{\alpha\beta\gamma\delta} v\rangle$ are the thermally averaged cross-sections for processes involving the annihilation of particles of sectors $\alpha\beta\rightarrow \gamma\delta$ ($\alpha,\beta,\gamma,\delta\in\{0,1,2\}$), given by \cite{Gondolo:1990dk,Alguero:2022inz}
    \begin{eqnarray}\label{eq:sigmavth}
    \langle \sigma_{\alpha\beta\gamma\delta} v\rangle=\frac{T}{8m_{\alpha}^2m_{\beta}^2K_{2}(\frac{m_\alpha}{T})K_{2}(\frac{m_\beta}{T})}\int_{(m_\alpha+m_\beta)^2}^\infty\sigma_{\alpha\beta\rightarrow\gamma\delta}(s)\big(s-(m_\alpha+m_\beta)^2\big)\sqrt{s}K_1\bigg(\frac{\sqrt{s}}{T}\bigg)ds.\nonumber
    \end{eqnarray}
In the last term on the right hand side of the Boltzmann equations, $\Gamma_{2\rightarrow 1}$ is the conversion term, which includes both the interaction rate of the co-scattering process and decay/inverse-decay and is given by
\begin{equation}
    \Gamma_{2\rightarrow1}=\sum_i\Gamma_{\alpha\rightarrow \beta,\rm SM}\frac{K_1(M_{\alpha}/T)}{K_2(M_{\alpha}/T)}+\langle \sigma_{2010} v \rangle n^{\rm eq}_{\rm SM},
\end{equation}
where $\alpha$ denotes sector 2 particles and $\beta$ is the DM. Here, the first term corresponds to the decays and inverse decays and the second term corresponds to the coscattering. We use \texttt{micrOMEGAs} \cite{Alguero:2022inz} to compute the relic density of dark matter.

\begin{figure}
    \centering
    \includegraphics[scale=0.45]{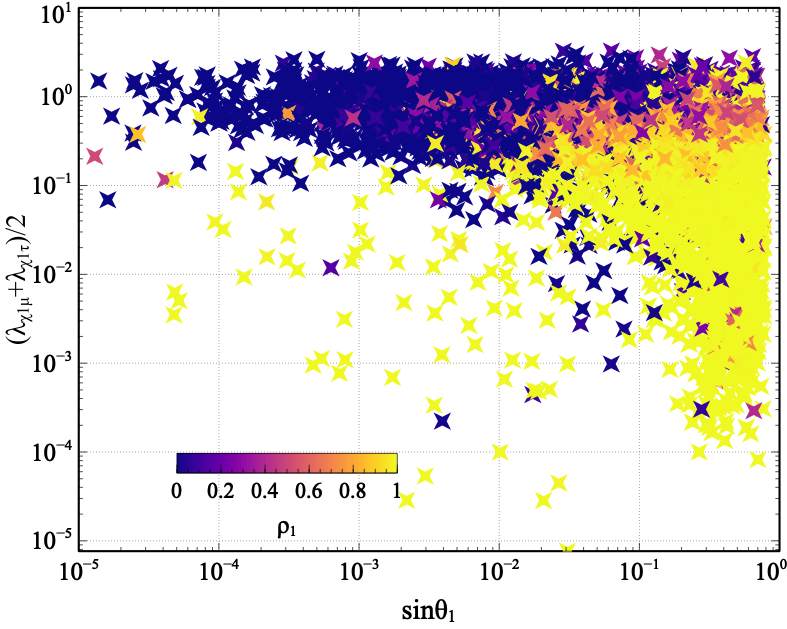}
    \caption{Parameter space consistent with correct DM relic density in the plane of $(\lambda_{\chi_{1\mu}}+\lambda_{\chi_{1\tau}})/2$ versus $\sin\theta_1$ with parameter $\rho_1$ in the color code with the condition $\rho_2\gtrsim0.9$.}
    \label{fig:correctrelic}
\end{figure}

To explain the DM relic density parameter space, we define two parameters $\rho_1$ and $\rho_2$ given as
\begin{eqnarray}
    \rho_1=\frac{\Omega_{DM}h^2}{\Omega'_{DM}h^2},\quad
    \rho_2=\frac{\Omega''_{DM}h^2}{\Omega_{DM}h^2}
    \label{eq:rhoparams}
\end{eqnarray}
where $\Omega_\text{DM}h^2$ represents the DM relic density and is estimated by including all the dark sector particles in the relic calculation, $\Omega'_\text{DM}h^2$ is defined as the DM relic density without the contribution of $\chi\chi\leftrightarrow\nu_R\nu_R$ channel, and $\Omega''_\text{DM}h^2$ is estimated without accounting for the contribution from heavier generation of singlet-doublet fermions. $\rho_1\simeq0$ implies that the DM relic density is dominantly decided by the DM self-annihilation to RHN channel. On the contrary, $\rho_1\simeq1$ implies that the DM relic density is independent of the $\chi\chi\leftrightarrow\nu_R\nu_R$ process. Similarly, $\rho_2\simeq1$ implies no contribution from heavier generation of SD fermions; on the other hand, any deviation from 1 implies the DM relic density dependence on heavier generation fermions. We now divide the DM parameter space into three regions:
\begin{enumerate}[label=(\roman*)]
\item Relic density decided by the process $\chi \chi \leftrightarrow \nu_R \nu_R$ (typically, $\rho_1<0.5$ and $\rho_2>0.8$). 
\item Relic density dominantly decided by DM annihilation to SM particles via Higgs and gauge processes along with its coannihilation with doublet fermion and singlet scalar $\phi$ to SM thermal bath (typically, $\rho_1>0.5$ and $\rho_2>0.8$). 
\item Relic density depends on the contribution from the decay of heavier generation singlet-doublet fermions ($\rho_2\lesssim0.8$).
\end{enumerate}

\noindent 
\textbf{Region (i):}
\begin{figure}
    \centering
    \includegraphics[scale=1.5]{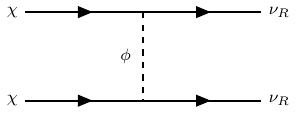}
    \caption{Feynman diagram showing the process $\chi\chi\rightarrow\nu_R\nu_R$ as a t-channel process}
    \label{fig:chichinuRnuR}
\end{figure}
\begin{figure}
    \centering
    \includegraphics[scale=0.45]{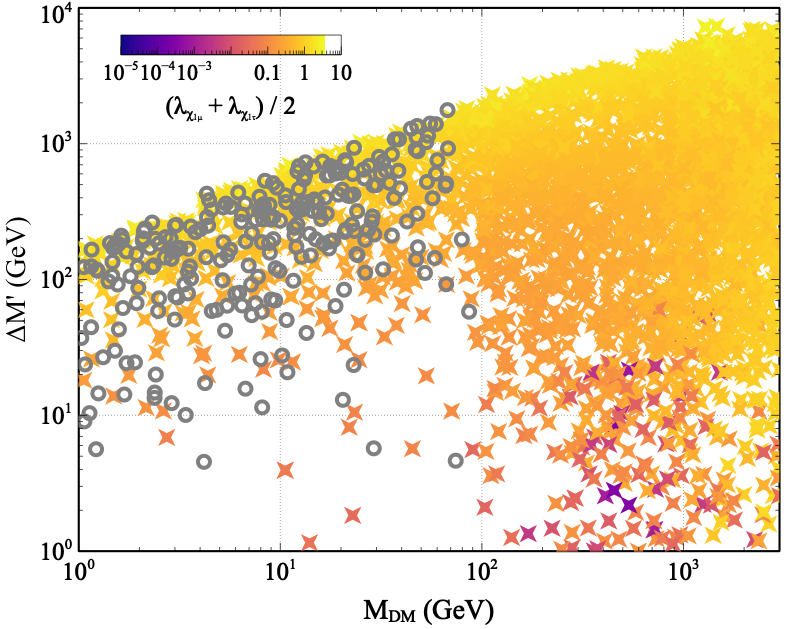}
    \caption{Correct relic density satisfying points with $\rho_1<0.5$ and $\rho_2>0.9$ are shown in the plane of $\Delta{M'}$ versus $M_{\rm DM}$ with $(\lambda_{\chi_{1\mu}}+\lambda_{\chi_{1\tau}})/2$ in the color code. The gray points are excluded from the LEP bound.}
    \label{fig:correctrelwithnuR}
\end{figure}

In this region, we choose the points from Fig.~\ref{fig:correctrelic} with $\rho_1\lesssim0.5$ and $\rho_2>0.8$, where the DM dominantly annihilates to RHNs via $\chi\chi\leftrightarrow \nu_R\nu_R$ processes, mediated by $\phi$ as shown in Fig.~\ref{fig:chichinuRnuR}.
In this scenario, we show the correct relic density satisfying points in the plane of $M_\text{DM}$ and $\Delta M'(\equiv M_{\phi_2}-M_\text{DM})$ in Fig.~\ref{fig:correctrelwithnuR}. The color band shows the value of the Yukawa coupling ($(\lambda_{\chi_{1\mu}}+\lambda_{\chi_{1\tau}})/2$).

\noindent
\textbf{Region (ii):}
\begin{figure}
    \centering
    \includegraphics[scale=0.45]{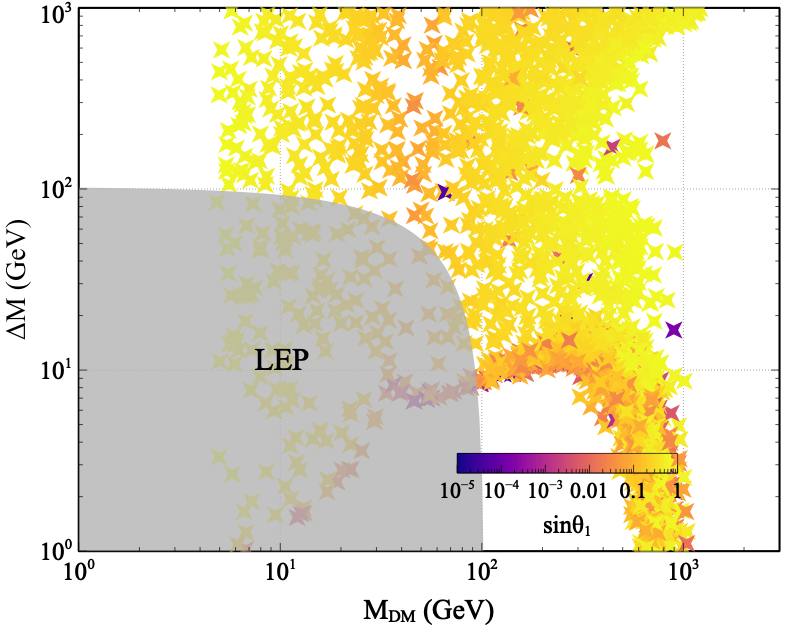}
    \caption{Correct relic density satisfying points with $\rho_1>0.5$ and $\rho_2>0.9$ are shown in the plane of $\Delta{M}$ versus $M_{\rm DM}$ with $\sin\theta_1$ in the color code.}
    \label{fig:correctrelpurelimit}
\end{figure}
In this region, we choose the points from Fig.~\ref{fig:correctrelic} with $\rho_1>0.5$ and $\rho_2>0.8$. The DM relic density is dominantly decided by DM annihilation to SM particles via Higgs and gauge mediated processes along with its coannihilation with doublet fermion and singlet scalar $\phi$ to SM thermal bath. The contribution from these processes is decided by the DM mass $M_\text{DM}$, the singlet-doublet mass splitting $\Delta{M}$, the singlet scalar DM mass splitting $\Delta M'$, the singlet-doublet mixing angle $\sin\theta_1$ and the Yukawa couplings $\lambda_{\psi}$. We showcase the correct relic density satisfying points in Fig.~\ref{fig:correctrelpurelimit} in the plane of $M_\text{DM}$ and $\Delta M$ and the color band represents the value of $\sin\theta_1$. The gray shaded region shows the disallowed region from LEP \cite{Barbieri:2004qk} bounds. For $\Delta M, \Delta M' \gtrsim 100$ GeV, the DM relic density is predominantly determined by DM self-annihilation into SM particles, which requires a relatively large mixing angle, typically $\sin\theta_1 \gtrsim 0.1$. On the other hand, for $\Delta M < 100$ GeV and/or $\Delta M' < 100$ GeV, the relic density is predominantly governed by co-annihilation of the DM with its doublet counterpart\footnote {We find co-scattering processes do not affect the DM relic density, as for $\sin\theta_1\gtrsim 0.05$, DM relic density is dominantly decided via $\chi\chi\leftrightarrow \nu_R\nu_R$.} and/or the scalar singlet $\phi$.\\

\begin{figure}
    \centering
    \includegraphics[scale=0.45]{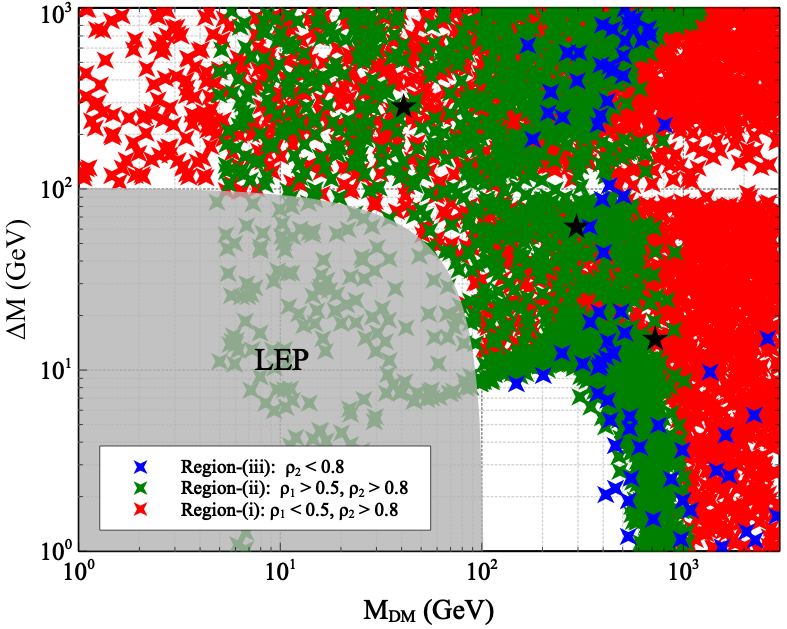}
    \caption{The correct relic density satisfying points in the plane of $\Delta{M}$ versus $M_{\rm DM}$ considering the dominant contributions from the three different regions in three colors as mentioned in the offset of the figure. Benchmark points from Table~\ref{tab:newBPs} are shown using black stars.}
    \label{fig:DMsummary}
\end{figure}

\noindent
\textbf{Region (iii):}
In the region we choose the correct relic density satisfying points with $\rho_2<0.8$, where the DM relic density becomes under-abundant because of large DM self-annihilation and/or singlet-doublet co-annihilation. The required relic density can still be achieved from the decay of heavier generation singlet-doublet fermions. We showcase these points in blue colored points in the plane of $\Delta M$ versus $M_\text{DM}$ in Fig.~\ref{fig:DMsummary}.

\subsection{DM Direct Detection}
\begin{figure}
    \centering
    \includegraphics[scale=1.2]{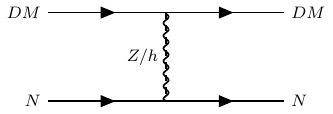}
    \centering
    \includegraphics[scale=1.2]{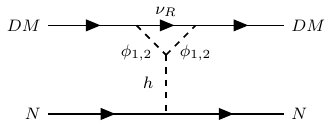}
    \caption{Spin-independent elastic DM-nucleon scattering arising at tree and one-loop level.}
    \label{fig:DD_feynman}
\end{figure}
In this scenario, owing to the singlet-doublet mixing, the DM $\chi$ can interact with the target nucleus in terrestrial direct search experiments through $Z$ and Higgs mediated processes. This cross-section for $Z$-boson mediated DM-nucleon scattering is \cite{Goodman:1984dc,Essig:2007az}
\begin{eqnarray}
    \label{eq:DDZmediated}
    \sigma_{SI}^Z=\frac{G_F^2\sin^4\theta_1}{\pi A^2}\mu_r^2\lvert\left[Z f_p+(A-Z)f_n\right]^2\rvert^2,
\end{eqnarray}
where $f_p=f_n=0.33$ correspond to the form factors for proton and neutrons, respectively with $\mu_r$ being the reduced mass of the DM-nucleon system, $A, Z$ being the mass number and the atomic number of the nucleus respectively. Similarly, the spin-independent DM-nucleon scattering cross-section through Higgs mediation is given by \cite{Cirelli:2013ufw}
\begin{eqnarray}
    \label{eq:DDHmediated}
    \sigma_{SI}^h=&&\frac{4}{\pi A^2}\mu_r^2\frac{y_1^2\sin^22\theta_1}{M_h^4}\left[\frac{m_p}{v}\left(f_{Tu}^p+f_{Td}^p+f_{Ts}^p+\frac{2}{9}f_{TG}^p\right)\right.\\
    &&\left.+\frac{m_n}{v}\left(f_{Tu}^n+f_{Td}^n+f_{Ts}^n+\frac{2}{9}f_{TG}^n\right)\right]^2,\nonumber
\end{eqnarray}
where $f_{Tu}^p=0.020\pm0.004$, $f_{Td}^p=0.026\pm0.005$, $f_{Ts}^p=0.118\pm0.062$, $f_{Tu}^n=0.014\pm0.003$, $f_{Td}^n=0.036\pm0.008$ and $f_{Ts}^n=0.118\pm0.062$ are the different coupling strengths between DM and light quarks in proton and neutron, respectively \cite{Bertone:2004pz,Ellis:2000ds}. The coupling of DM with the gluons in the target nuclei are parameterized by \cite{Hoferichter:2017olk} $f_{TG}^{p,n}=1-\sum_{q=u,d,s}f_{Tq}^{p,n}$. Beside these tree-level processes for DM-nucleon scattering, there exists another contribution to the spin-independent direct search cross-section, which arises at the loop level with $\nu_{R_\beta}$ and $\phi_{1,2}$ in the loop. The corresponding Feynman diagram is as shown in Fig.~\ref{fig:DD_feynman}. This cross section can be evaluated as \cite{Ibarra:2016dlb}:
\begin{eqnarray}
    \label{eq:DDloopmediated}
    \sigma_{SI}^{\rm{loop}}=\frac{4}{\pi}\frac{M_{\chi}^2m_p^2}{(M_\chi+m_p)^2}m_p^2\,\zeta^2f_q^2
\end{eqnarray}
where the form factor $f_q\approx0.3$ and the loop-induced effective coupling $\zeta$ is given by,
\begin{eqnarray}
    \label{eq:DDloopfactor}
    \zeta=\sum_\beta\frac{\lambda_{\chi_{1\beta}}^2}{16\pi^2M_h^2M_{\chi}^2}\lambda_{\phi H}\,F\left(\frac{M_\chi^2}{M_\phi^2}\right)
\end{eqnarray}
where the loop function $F(x)$ is given by
\begin{eqnarray}
    \label{eq:DDloopfunc}
    F(x)=\frac{x+(1-x)\ln(1-x)}{x}.
\end{eqnarray}
\begin{figure}
    \centering
    \includegraphics[scale=0.45]{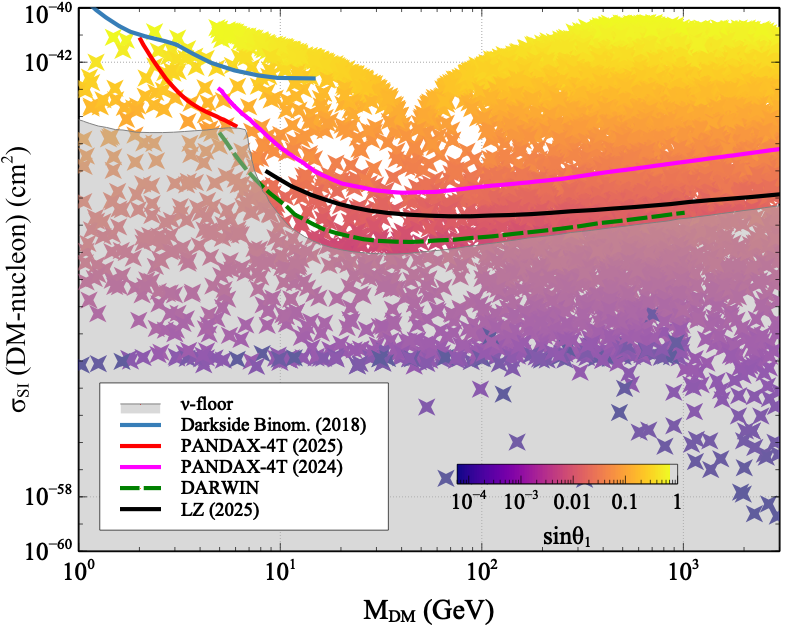}
    \caption{Spin independent direct detection cross-section as a function of DM mass. }
    \label{fig:DD}
\end{figure}
In this scenario, spin-independent (SI) DM direct detection is possible with the SM Higgs and $Z$ Boson exchange diagrams. We calculate the SI direct-detection cross-sections for the points shown in Fig.~\ref{fig:correctrelic} and impose constraints from LZ \cite{LZ:2024zvo}, PANDAX-4T (2024) \cite{PandaX:2024qfu}, PANDAX-4T (2025) \cite{PandaX:2025rrz} and Darkside binom. \cite{DarkSide:2018bpj}. Fig.~\ref{fig:DD} shows corresponding points in the plane of SI DM-nucleon cross-section $\sigma_{SI}$ and DM mass with mixing angle $\sin\theta_1$ in the color code.  The solid contours correspond to different experimental constraints while the green dashed contour shows the sensitivity of future experiment DARWIN~\cite{DARWIN:2016hyl}. The gray shaded region denotes the neutrino floor \cite{Billard:2021uyg} where coherent neutrino-nucleus scattering acts as a background to any DM-induced event.

\section{Inverse Phase Transition}
\label{sec:InvPT}
The possibility of a first-order EWPT in SD dark matter model with radiative neutrino mass was studied earlier \cite{Dey:2025pcs} where the singlet scalars play the dominant role in generating the desired effective potential at finite-temperature. In the present work, we consider SD fermion driven inverse phase transition and hence ignore the contribution to the effective potential from singlet scalars. Therefore, the tree level contribution to the effective scalar potential arises from the SM Higgs sector alone and is expressed as
\begin{equation}
V_0(h)=\frac{m_h^2}{8v^2}\left(h^2-v^2\right)^2.
\end{equation}
To investigate the thermal evolution of the effective potential in the early Universe, we construct the one-loop effective scalar potential at finite temperature by including the contributions from all SM particles that couple significantly to the Higgs field, namely the $W$ and $Z$ gauge bosons, the top quark, the Higgs boson, the Goldstone bosons, together with two generations of singlet-doublet fermions. The field-dependent masses of these particles are denoted by $m_i(h)$, where $h$ is the background Higgs field and the index $i$ labels the corresponding particle species. For the relevant SM particles, the field-dependent masses are given by
\begin{align}
m_{W,Z}(h)^2 &= \frac{m_{W,Z}^2}{v^2}h^2, \\
m_t(h)^2 &= \frac{m_t^2}{v^2}h^2, \\
m_h(h)^2 &= \frac{m_h^2}{2v^2}\left(3h^2-v^2\right), \\
m_g(h)^2 &= \frac{m_h^2}{2v^2}\left(h^2-v^2\right).
\end{align}
After diagonalizing the singlet-doublet fermion mass matrix, the field-dependent mass eigenvalues for the two generations are given by 

\begin{align}
    m_{X}(h)^2 &=\frac{1}{4}\left(M_{\chi_i}+M_{\Psi_i}-\sqrt{(M_{\chi_i}-M_{\Psi_i})^2+4h^2y_i^2}\right)^2\\
    m_{Y}(h)^2 &=\frac{1}{4}\left(M_{\chi_i}+M_{\Psi_i}+\sqrt{(M_{\chi_i}-M_{\Psi_i})^2+4h^2y_i^2}\right)^2,
\end{align}
where $X\in\{\chi,\chi'\}$, $Y\in\{\psi,\psi'\}$, and $i\in\{1,2\}$ represents fermion generation index. The zero-temperature one-loop correction to the scalar potential, commonly referred to as the Coleman Weinberg (CW) potential~\cite{Coleman:1973jx}, receives contributions from the SM particles as well as the additional singlet-doublet fermions. It is given by
\begin{align}
    V_{\rm CW}=\frac{1}{64\pi^2}\sum_{i=W,Z,h,g,t,X,Y}\left(n_i m_i(h)^4\left(\log\left(\frac{m_i(h)^2}{v^2}\right)-C_i\right)\right),
\end{align}
where $n_{W,Z,h,g,t,X,Y}=\{6,3,1,3,-12,-4,-4\}$ and \(C_i\) denotes the renormalization scheme dependent constant, taking the value \(C_i=5/6\) for gauge bosons and \(C_i=3/2\) for all other particle species. At finite temperature, the effective potential receives additional thermal corrections arising from the interactions of the Higgs field with the thermal plasma~\cite{Dolan:1973qd,Quiros:1999jp}. The finite-temperature contribution to the effective potential is given by
\begin{equation}
V_T(h,T)=\frac{T^4}{2\pi^2}\sum_i n_i
J_{B/F}\!\left(\frac{m_i(h)^2}{T^2}\right),
\end{equation}
where \(J_B(y)\) and \(J_F(y)\) denote the bosonic and fermionic thermal functions, respectively, defined as
\begin{equation}
J_{B,F}(y)=\int_0^\infty dx\,x^2
\ln\!\left(1\mp e^{-\sqrt{x^2+y}}\right).
\end{equation}
To account for the infrared (IR) divergences arising from bosonic zero modes at high temperatures, we include the daisy (ring) resummation~\cite{Fendley:1987ef,Parwani:1991gq,Arnold:1992rz} contribution to the effective potential following the Arnold Espinosa method~\cite{Arnold:1992rz}. The corresponding correction is given by
\begin{equation}
V_{\rm daisy}(\phi,T)
=
-\sum_{i=W,Z,\gamma,h,g}
\frac{\bar{n}_i T}{12\pi}
\left[
\left(m_i^2(\phi)+\Pi_i(T)\right)^{3/2}
-
m_i^3(\phi)
\right],
\end{equation}
where the effective numbers of degrees of freedom are $\bar{n}_{W,Z,\gamma,h,g}=\{2,1,1,1,3\}$.
The thermal (Debye) masses entering the daisy correction are given by
\begin{align}
\Pi_{h,g}(T)
&=
\frac{T^2}{4v^2}
\left(
m_h^2+m_Z^2+2m_W^2+2m_t^2
\right),\\
\Pi_W(T)
&=
\frac{22\,T^2}{3v^2}m_W^2,\\
\Pi_Z(T)
&=
\frac{22\,T^2}{3v^2}
\left(m_Z^2-m_W^2\right)
-
m_W^2(h),\\
\Pi_\gamma(T)
&=
m_W^2(h)
+
\frac{22\,T^2}{3v^2}m_W^2.
\end{align}
The full finite-temperature effective potential used to study the inverse first-order electroweak phase transition is given by
\begin{equation}
V_{\rm eff}(h,T)
=
V_0(h)
+
V_{\rm CW}(h)
+
V_T(h,T)
+
V_{\rm daisy}(h,T).
\end{equation}
In this model, the Universe undergoes a distinctive thermal history during its evolution in contrast with typical scenarios with a first-order phase transition where the Universe goes from a symmetric to a broken phase at one step. At very high temperatures, well above the electroweak scale, the electroweak symmetry is broken through a crossover transition, and the Higgs field acquires a nonzero vacuum expectation value. As the temperature decreases, the vacuum with nonzero VEV becomes metastable, while the symmetric phase with a vanishing VEV emerges as the global minimum of the effective potential. Consequently, the Universe undergoes a first-order phase transition from the broken phase to the symmetric phase at a temperature above the electroweak scale. Upon further cooling to temperatures around the electroweak scale, the symmetric vacuum again becomes metastable, and the Universe undergoes a second first-order phase transition to the electroweak symmetry breaking vacuum with a nonzero VEV. 

\begin{figure}
    \centering
    \includegraphics[width=\linewidth]{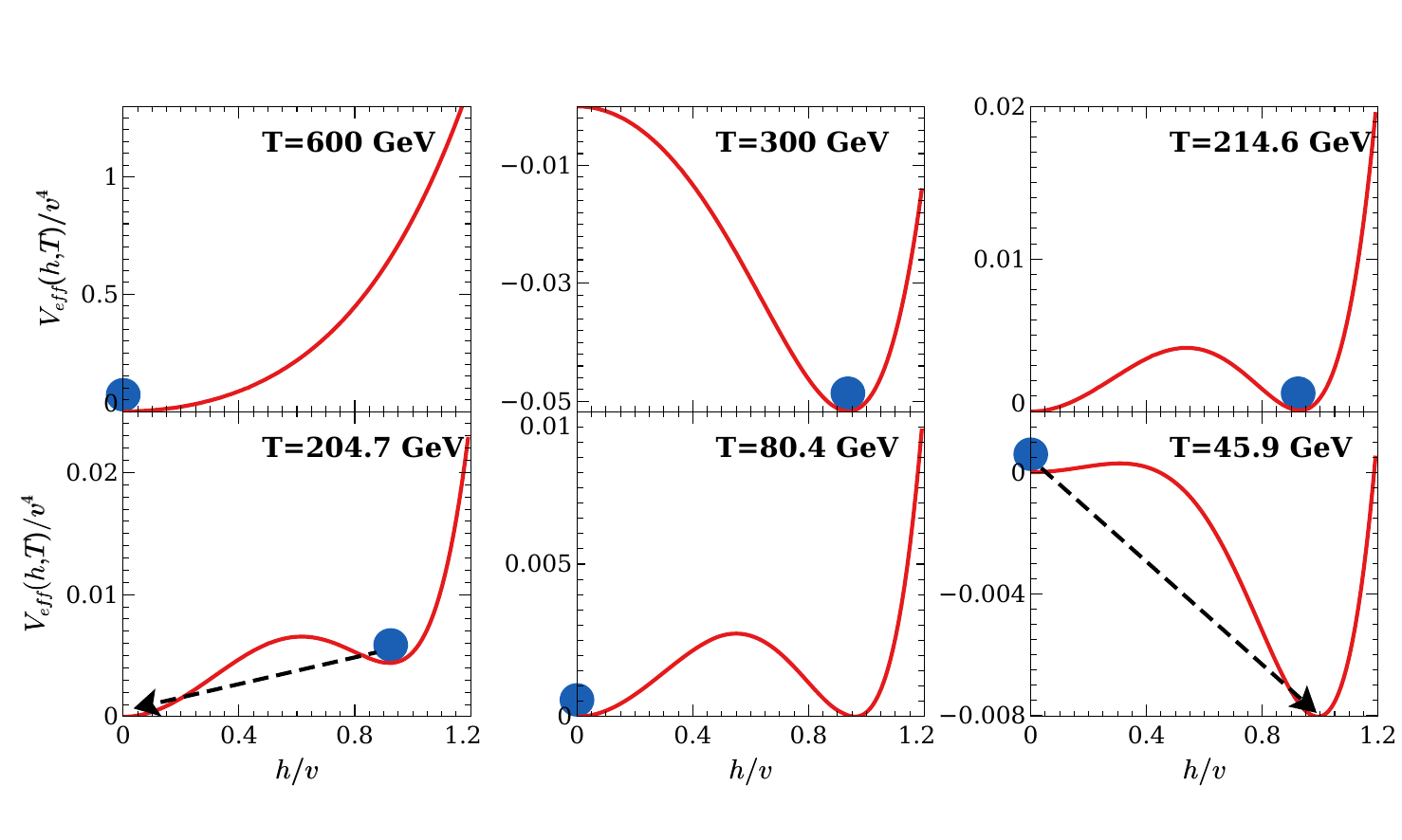}
    \caption{Scalar potential is plotted against $h$ field scaled with Higgs vev $v$ for BP1.}
    \label{fig:potential_BP1}
\end{figure}

The thermal evolution of the effective potential is illustrated in Fig.~\ref{fig:potential_BP1} for a benchmark point BP1 given in Table \ref{tab:newBPs}. At $T=600~\mathrm{GeV}$, the Universe resides in the symmetric phase, corresponding to the vacuum with a vanishing VEV, indicated by the blue dot. As the temperature decreases below $300~\mathrm{GeV}$, a second minimum with a nonzero VEV develops and becomes the preferred vacuum. Upon further cooling, at $T_{c1}=214.6~\mathrm{GeV}$, the symmetric and broken minima become degenerate in energy, marking the first critical temperature of the phase transition. This degeneracy signals the onset of a first-order phase transition from the broken phase to the symmetric phase. Below the first critical temperature, the symmetric and broken minima are separated by a potential barrier, allowing the phase transition to proceed through quantum tunneling. The tunneling rate per unit volume is given by $\Gamma(T)=T^4\left(\frac{S_3}{2\pi T}\right)^{3/2}e^{-S_3/T}$,
where $S_3$ is the three-dimensional Euclidean bounce action~\cite{Linde:1980tt}. The nucleation temperature, $T_{n1}$, is defined by the condition $\Gamma(T_{n1})=\mathcal{H}^4(T_{n1})$, with $\mathcal{H}$ denoting the Hubble expansion rate. At this temperature, bubbles of the symmetric phase nucleate and expand, causing the Universe to transition from the broken phase to the symmetric phase.
As the Universe cools further, the effective potential evolves again, and at $T_{c2}=80.4~\mathrm{GeV}$ the symmetric and broken minima become degenerate for the second time, defining the second critical temperature. Below $T_{c2}$, the broken minimum becomes the true vacuum, and the Universe undergoes a second first-order phase transition at the corresponding nucleation temperature, $T_{n2}$, through the nucleation and expansion of true-vacuum bubbles.

\begin{figure}
    \centering
    \includegraphics[scale=0.45]{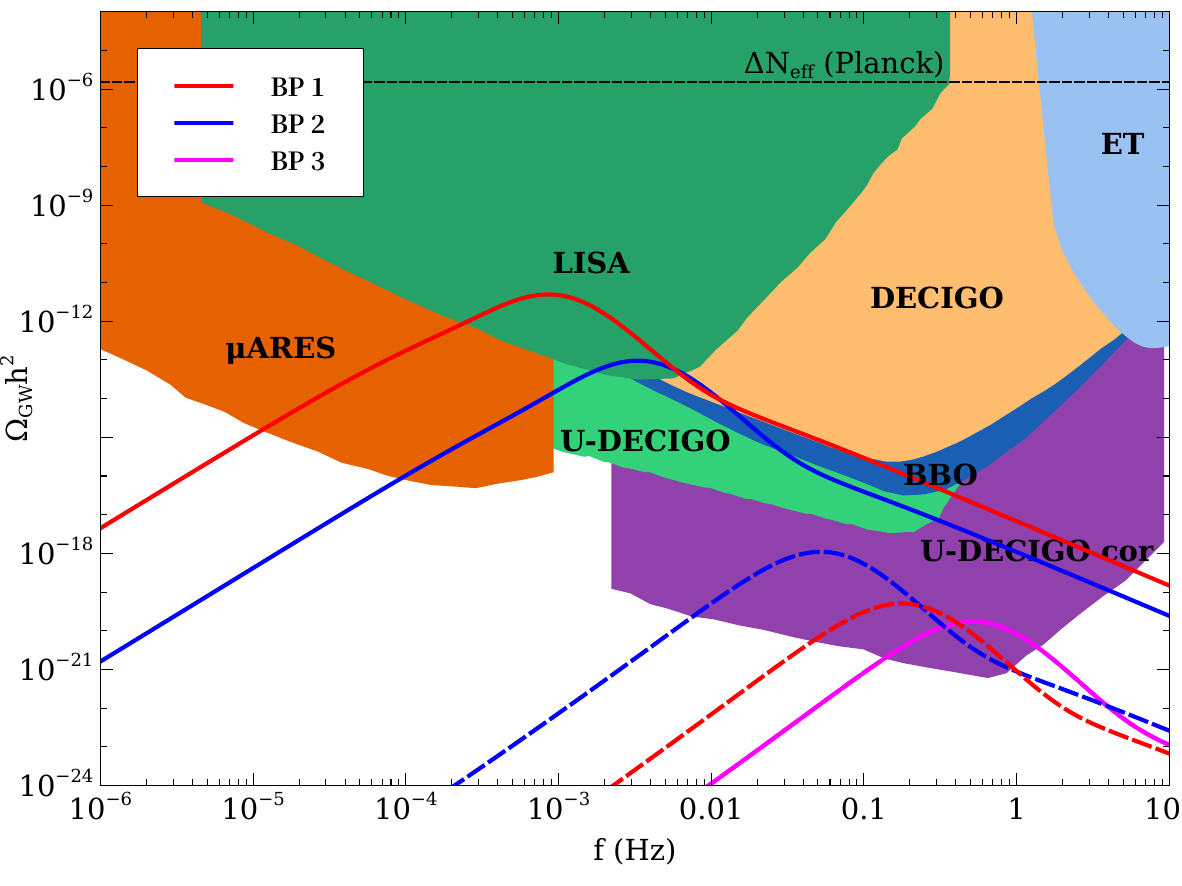}
    \caption{GW spectrum corresponding to the benchmark points given in Table \ref{tab:newBPs}.}
    \label{fig:gw}
\end{figure}

In both the first-order phase transitions of our IFOEWPT scenario, the nucleation, expansion, and collision of true-vacuum bubbles generate a stochastic GW background. The resulting GW spectrum receives contributions from three primary sources: bubble wall collisions~\cite{Turner:1990rc,Kosowsky:1991ua,Kosowsky:1992rz,Kosowsky:1992vn,Turner:1992tz}, long-lasting sound waves in the plasma~\cite{Hindmarsh:2013xza,Giblin:2014qia,Hindmarsh:2015qta,Hindmarsh:2017gnf}, and magneto-hydrodynamic turbulence in the plasma~\cite{Kamionkowski:1993fg,Kosowsky:2001xp,Caprini:2006jb,Gogoberidze:2007an,Caprini:2009yp,Niksa:2018ofa}. The properties of the stochastic GW signal are primarily characterized by three phase transition parameters. The first one is the nucleation temperature, $T_n$, at which bubbles of the true vacuum are nucleated at an appreciable rate. The second one is the strength parameter,
$\alpha=\frac{\Delta\rho}{\rho_{\rm rad}}$,
defined as the ratio of the released vacuum energy (latent heat) to the radiation energy density at the nucleation temperature, where
$\Delta\rho=\Delta V_{\rm eff}-T\frac{\partial \Delta V_{\rm eff}}{\partial T},
\rho_{\rm rad}=\frac{\pi^2}{30}g_*T_n^4$,
with $\Delta V_{\rm eff}$ denoting the difference in the effective potential between the false and true vacua, and $g_*$ the effective number of relativistic degrees of freedom. The third parameter is the inverse duration parameter,
$\frac{\beta}{\mathcal{H}_n}=T_n\left.\frac{d}{dT}\left(\frac{S_3}{T}\right)\right|_{T=T_n}$,
where $\beta^{-1}$ represents the characteristic timescale of the phase transition and $\mathcal{H}_n$ is the Hubble parameter at the nucleation temperature. Physically, a larger value of $\alpha$ corresponds to a stronger first-order phase transition, while a smaller value of $\beta/\mathcal{H}_n$ indicates a longer-lasting phase transition. For the numerical evaluation of the three-dimensional Euclidean bounce action, $S_3$, we use the package \texttt{FindBounce}~\cite{Guada:2020xnz}. Fig.~\ref{fig:gw} shows the gravitational-wave spectra corresponding to the benchmark points listed in Tables~\ref{tab:newBPs} and \ref{tab:newBPsgw}. For each benchmark point, the dashed and solid curves represent the gravitational-wave signals from the inverse FOPT and the subsequent FOPT, respectively. Also shown are the projected sensitivities of the future gravitational-wave observatories $\mu$ARES \cite{Sesana:2019vho}, LISA \cite{2017arXiv170200786A}, DECIGO \cite{Kawamura:2006up}, UDECIGO (UDECIGO-corr) \cite{Sato:2017dkf, Ishikawa:2020hlo}, BBO \cite{Yagi:2011wg}, and ET \cite{Punturo_2010}, indicated by shaded regions of different colors. The horizontal dashed black dashed line denotes the upper bound on the stochastic GW background from the Big Bang Nucleosynthesis (BBN) constraint on the effective number of relativistic degrees of freedom~\cite{Planck:2018jri}.

\begin{table}
\centering
\setlength{\tabcolsep}{10pt}
\begin{tabular}{|l||c|c|c|c|c|c|c|}
\hline
\textbf{} & $\mathbf{M_{\chi}}$ & $\mathbf{\Delta M_{1}}$& $\mathbf{M_{\chi'}}$ & $\mathbf{M_{\psi'}}$ & $\mathbf{\sin\theta_{1}}$ & $y_2$ & $\mathbf{M_\phi}$\\
\hline
\textbf{BP1} & $40.96$  & $284.46$  & $400$ & $1661$ & $0.00407$ & $2.763$ & $189.10$ \\
\hline
\textbf{BP2} & $294.84$  &  $61.70$  & $600$ & $1965$ & $0.01264$ & $2.991$ & $594.12$ \\
\hline
\textbf{BP3} &  $724.59$  & $14.82$  & $800$ & $2225$ & $0.00972$ & $3.123$ & $783.01$ \\
\hline

\end{tabular}
\caption{Benchmark Points satisfy DM relic and inverse FOPT scenario.}
\label{tab:newBPs}
\end{table}

\begin{table}
\centering
\setlength{\tabcolsep}{10pt}
\begin{tabular}{|l||c|c|c|c|c|c|c|c|}
\hline
\textbf{} & $T_{c1}$ & $T_{n1}$& $\alpha_1$ & $\beta_1/\mathcal{H}$ & $T_{c2}$ & $T_{n2}$ & $\alpha_2$ & $\beta_2/\mathcal{H}$\\
\hline
\textbf{BP1} & $214.6$  & $204.72$  & $0.0015$ & $5916$ & $80.39$ & $45.93$ & $0.25$ & $180$\\
\hline
\textbf{BP2} & $245.4$  & $223.56$  & $0.002$ & $1606$ & $80.96$ & $56.89$ & $0.10$ & $510$\\
\hline
\textbf{BP3} & $283.0$  & $276.83$  & $0.0018$ & $12798$ & $88.70$ & $75.61$ & $0.03$ & $1358$\\
\hline

\hline
\end{tabular}
\caption{Parameters associated with the two FOPTs for the model parameters given in Table~\ref{tab:newBPs}.}
\label{tab:newBPsgw}
\end{table}

\section{Contribution of Dirac neutrinos to $\Delta N_{\rm{eff}}$}
\label{sec:Neff}
We now discuss the phenomenological implications arising from the presence of the right-handed counterparts of the SM neutrinos. Since neutrinos are Dirac particles in our framework, the right-handed neutrinos must be as light as the SM left-handed neutrinos. The existence of these additional ultra-light particles in the early Universe can increase the total radiation energy density and consequently modify the effective number of relativistic degrees of freedom, ($N_{\rm eff}$), which is defined as
\begin{eqnarray}
    \label{eq:Neffdefination}
    N_{\rm{eff}}=\frac{\rho_{\rm{rad}}-\rho_{\gamma}}{\rho_{\nu_L}},
\end{eqnarray}
where $\rho_{\rm rad}$ is the total radiation energy density, while $\rho_{\gamma}$, $\rho_{\nu_L}$ denote the energy densities of photons and a single active neutrino species, respectively. In the absence of any additional light degrees of freedom, the SM predicts $N_{\rm eff}=3.045$\footnote{The small deviation from 3 arises from non-instantaneous neutrino decoupling, neutrino flavour oscillations, and finite-temperature QED corrections to the electromagnetic plasma.}\cite{Mangano:2005cc,Grohs:2015tfy,deSalas:2016ztq,Cielo:2023bqp, Akita:2020szl, Froustey:2020mcq, Bennett:2020zkv}. The latest measurements of the cosmic microwave background (CMB) by the Planck satellite \cite{Planck:2018vyg}, combined with baryon acoustic oscillation (BAO) data, give $N_{\rm eff}=2.99^{+0.34}_{-0.33}$ at $95\%$ confidence level, which is in excellent agreement with the SM prediction. Recently, the combination of DESI BAO with CMB data \cite{Elbers:2025vlz} 
within the one-parameter extension $\Lambda$CDM+$N_{\rm eff}$ has provided a slightly weaker constraint $N_{\rm eff}=3.23^{+0.35}_{-0.34} \quad (95\%).$

Future experiments such as CMB-S4 \cite{Abazajian:2019eic}, SPT-3G \cite{SPT-3G:2019sok}, CMB-HD~\cite{CMB-HD:2022bsz} and Euclid \cite{Euclid:2024imf} are expected to measure $N_{\rm eff}$ with much higher precision, making it a powerful probe of physics beyond the SM.
The contribution of the light right-handed neutrinos to $\Delta N_{\rm eff}$ depends on their production history, namely whether they were thermalized in the early Universe or produced non-thermally during its evolution \cite{Luo:2020sho,Luo:2020fdt,Abazajian:2019oqj}. The interplay between the Dirac nature of neutrinos, dark matter production, and $\Delta N_{\rm eff}$ has been explored previously in Refs.~\cite{Biswas:2021kio,Biswas:2022vkq}. In our model, the right-handed neutrinos interact only through the Yukawa interaction $\lambda_{\chi}\overline{\nu_R}\phi\chi$, as shown in Eq.~\eqref{eq:lag}. Therefore, their production, whether thermal or non-thermal, is entirely controlled by the Yukawa coupling $\lambda_{\chi}$.

\begin{figure}
		\centering
		\includegraphics[scale=0.09]{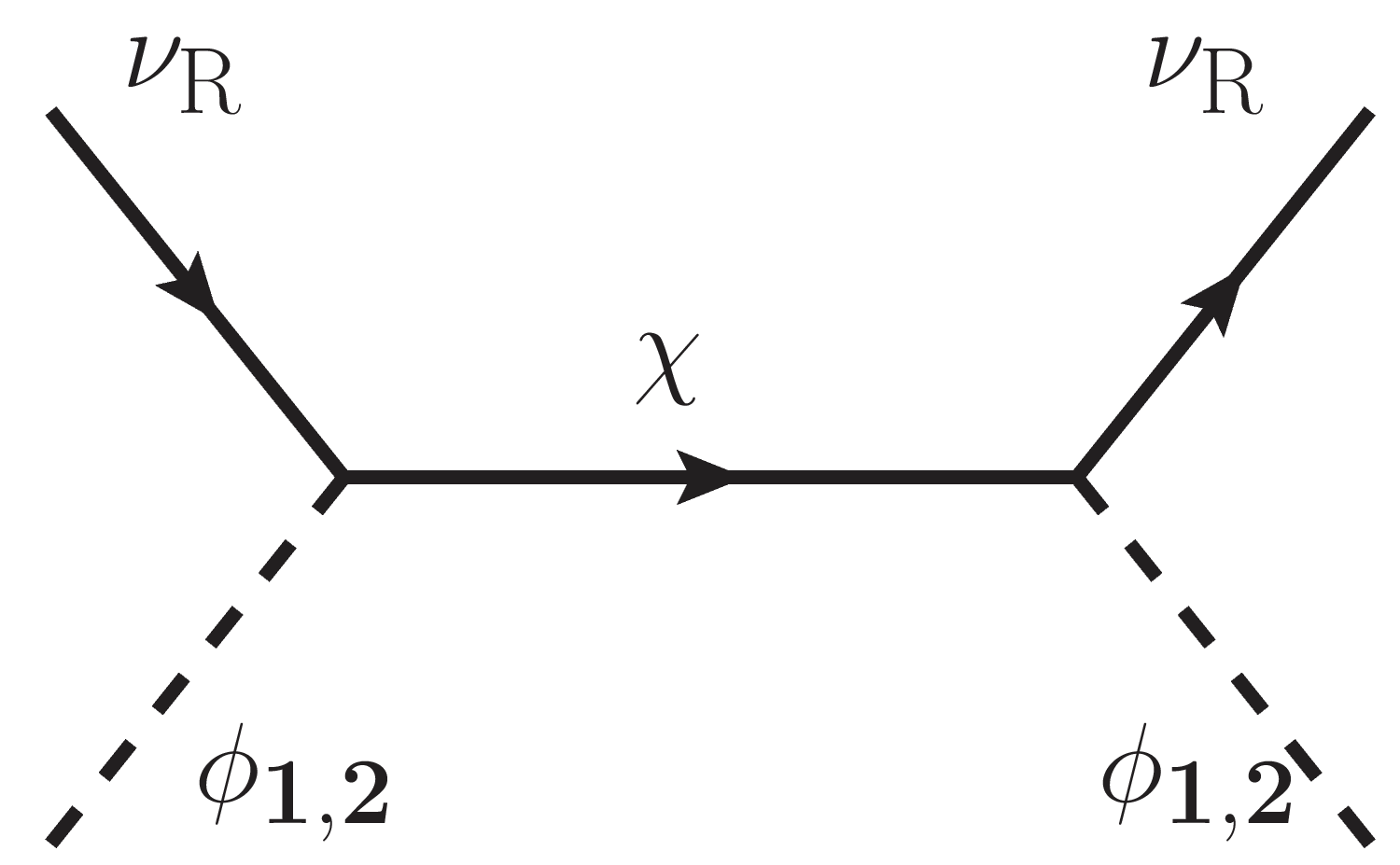}
		\includegraphics[scale=0.09]{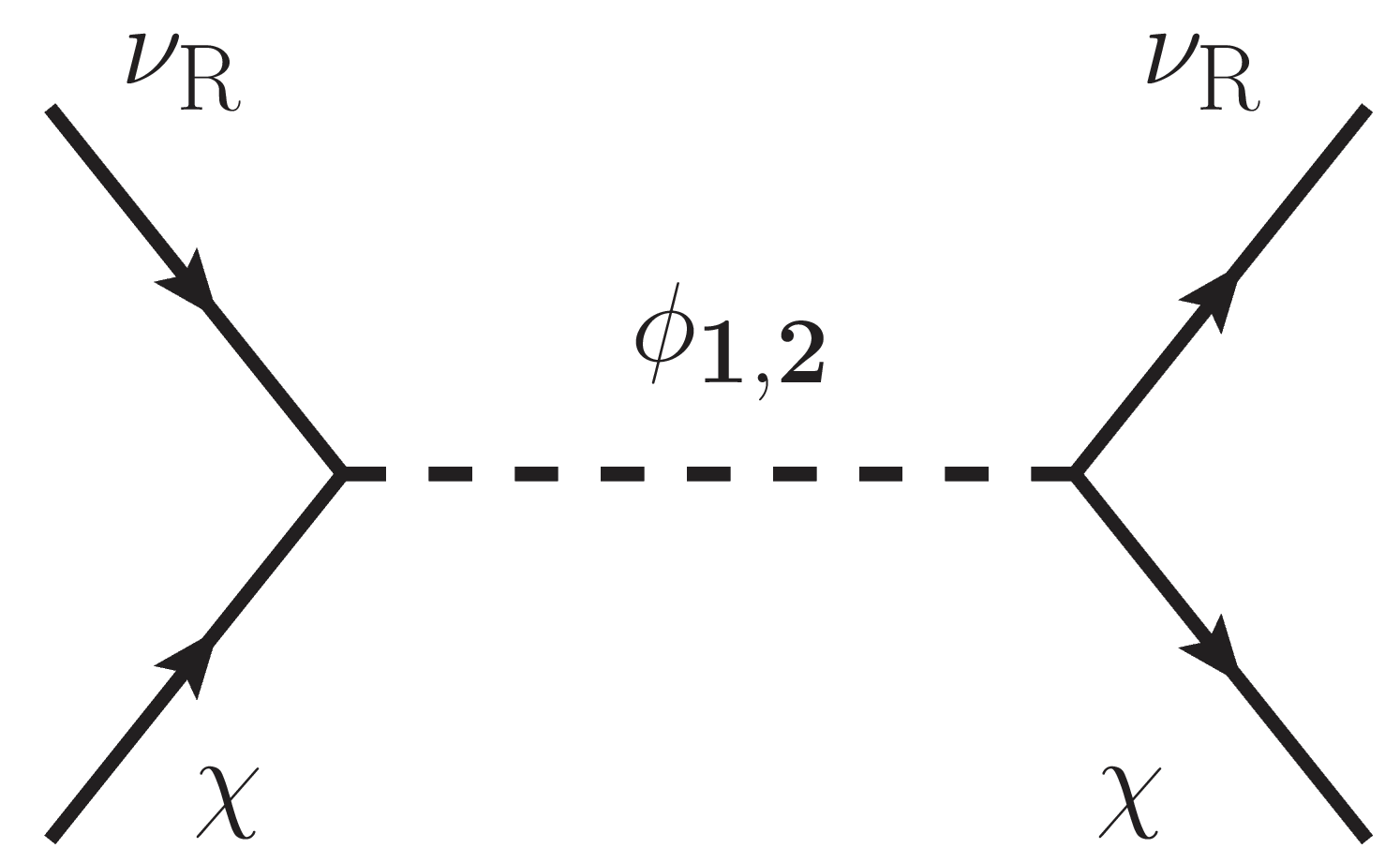}
		\caption{Feynman diagrams of the processes which keep $\nu_R$ in thermal equilibrium with the SM.}
		\label{fig:image4}
	\end{figure}

From Eq.~\eqref{eq:Neffdefination}, the additional contribution to $N_{\rm eff}$ at the time of CMB coming from the presence of $\nu_R$ in the total radiation energy density can be written as,
\begin{equation}
	\Delta{N_{\rm eff}}= N_{\nu_R}\times \frac{\rho_{\nu_R}}{\rho_{\nu_L}} \Bigg|_{\rm T=T_{\rm CMB}},
	\end{equation}
	where $N_{\nu_R}$ is the number of relativistic $\nu_R $ species, and $\rho_{\nu_R}$ is the energy density of the single $\nu_R$. In the equation above, we assume that muon-type and tau-type $\nu_R$ behave the same way. We also assume that the energy density of electron-type $\nu_R$ is essentially zero, since $\lambda_{\chi_{1e}} = \lambda_{\chi_{2e}} = 0$. Because of this, we take $N_{\nu_R}$ to be 2. Since, in our setup, the coupling $\lambda_{\chi}$ is large, the $\nu_R$ can be thermally produced in the early Universe from the annihilation of $\chi$ and $\phi$. Fig.~\ref{fig:image4} shows the interactions responsible for keeping $\nu_R$ in equilibrium. Once these interaction rates drop below the expansion rate of the Universe, $\nu_R$'s get decoupled relativistically from the thermal plasma and their temperature evolves independently. Hence by using entropy conservation in both sectors, one can write the excess contribution to radiation from $\nu_R$ as 
	\begin{equation}\label{eq:7c}
	\Delta N_{\rm eff}=N_{\nu_R}\times\left(\frac{T_{\nu_R}}{T_{\nu_L}}\right)^4=N_{\nu_R}\left(\frac{g_{*s}(T^{\rm dec}_{\nu_L})}{g_{*s}(T^{\rm dec}_{\nu_R})}\right)^{4/3},
	\end{equation} 
	where $g_{*s}(T_{\kappa}^{\rm dec})$ is the relativistic entropy degrees of freedom at temperature $T_{\kappa}^{\rm dec}$ which is the decoupling temperature of the species $\kappa(\equiv \nu_L, \nu_R )$ from the thermal bath. The Yukawa coupling $\lambda_{\chi}$ which leads to thermalization of $\nu_R$ also plays an important role in neutrino mass generation. Increasing $\lambda_{\chi}$ corresponds to a smaller mixing angle $\sin\theta$ to satisfy the correct neutrino mass. The viable parameter space for $\Delta N_{\rm eff}$ is obtained by satisfying all possible constraints such as neutrino mass, $(g-2)_\mu$ and DM relic density as well as direct-detection rate.
\begin{figure}
    \centering
    \includegraphics[scale=0.45]{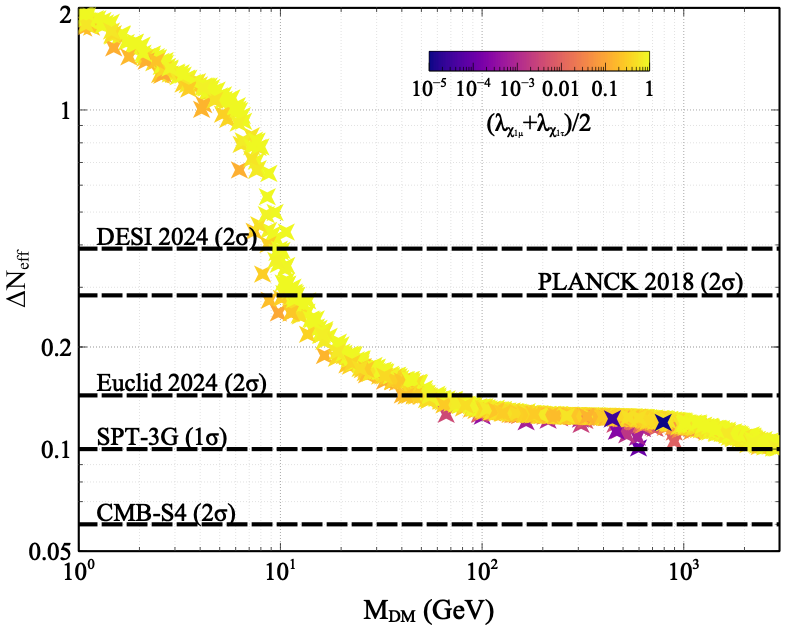}
    \caption{Variation of $\Delta{N_{\rm{eff}}}$ as a function of DM mass $M_{\rm{DM}}$ with $\lambda_{\chi_1}$ in the color code.}
    \label{fig:Neff}
\end{figure}
Fig.~\ref{fig:Neff} shows the variation of $\Delta N_{\rm eff}$ as a function of DM mass ranging from 1 GeV to 1 TeV. The color gradient shows the dependence on $\lambda_\chi$. One can clearly see a correlation among $\Delta{N_{\rm eff}}$, $\lambda_{\chi}$ and the mass of DM. Larger $\lambda_{\chi}$ and lighter $M_{\rm DM}$ corresponds to larger contribution to the $\Delta{N_{\rm eff}}$. This can be understood as follows. A larger $\lambda_{\chi}$ or lighter $M_{\rm DM}$ forces $\nu_R$ to be in the thermal bath for a longer time. The late decoupling of $\nu_R$ from the plasma decreases the denominator in Eq.~\eqref{eq:7c} and increases its contribution to the $N_{\rm eff}$.

\section{Collider Signatures}
\label{sec:Collider}
\begin{figure}
    \centering
    \includegraphics[scale=0.45]{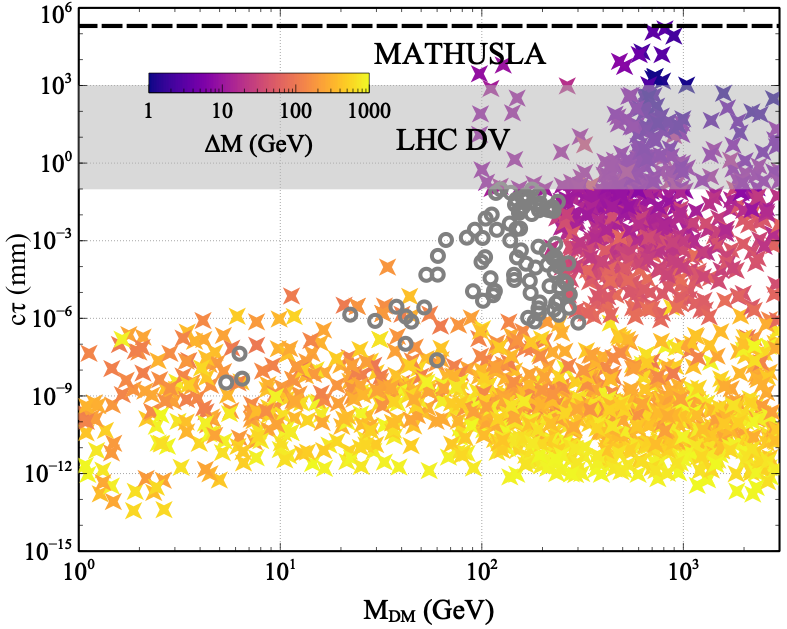}
    \caption{Decay length of the doublet fermion as a function of DM mass with $\Delta{M}$ in the color code. The gray points in the plot are ruled out by prompt decay bounds from ATLAS.}
    \label{fig:collidersign}
\end{figure}
In this section, we briefly discuss the collider signatures of the model. The first generation of charged doublet fermion, being in the sub-TeV range, $\psi_1^\pm$ can be abundantly produced via gauge interactions and subsequently decay into DM and charged leptons through singlet–doublet mixing, mediated by off-shell gauge bosons \cite{Bhattiprolu:2025beq,Thomas:1998wy,Djouadi:2001fa}. The collider signature depends on its decay length, which is controlled by the mixing angle $\sin\theta_1$. For small mixing \cite{Paul:2024prs,Paul:2025spm}, the decay width is suppressed, leading to long-lived charged particles that can give rise to displaced vertex signatures at the LHC \cite{ATLAS:2022gbw,CMS:2024trg} and at future detectors such as MATHUSLA \cite{MATHUSLA:2019qpy}. On the other hand, when the decay length is smaller than $0.1$ mm \cite{Paul:2024prs,Paul:2025spm}, the decay becomes prompt before reaching the detector volume, resulting in standard collider signatures with missing transverse energy and multilepton final states, such as $pp \rightarrow 3\ell + E_T^{\rm miss}$\cite{CMS:2021cox,ATLAS:2021moa}. These complementary signatures provide important probes of the model across different regions of parameter space \cite{Paul:2024prs,Paul:2025spm}.
In Fig.~\ref{fig:collidersign}, we show the decay length of $\psi_1^{\pm}$ ($c\tau$ in mm) as a function of DM mass with $\Delta{M}$ in the color code. The gray points in the plot are excluded from the bounds on prompt decays set by CMS and ATLAS \cite{CMS:2021cox,ATLAS:2021moa}. The sensitivity of LHC DV is shown by the light gray shaded region and the MATHUSLA projection is shown by black dashed line. We note that, the decay width increases with the increase in $\Delta{M}$ as given in Appendix~\ref{app:decayrate}. Hence the region with small $\Delta{M}$ results in large $c\tau$ and falls within and above the sensitivity region of LHC DV, which is presented in Fig.~\ref{fig:collidersign}.

\section{Summary and Conclusion}
\label{sec:results}
\begin{figure}
    \centering
    \includegraphics[scale=0.45]{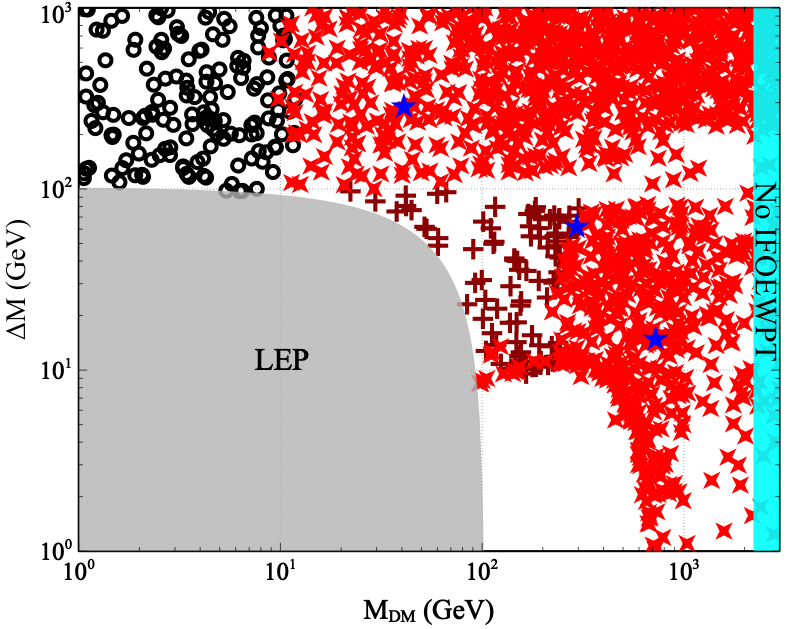}
    \caption{Summary of allowed parameter space satisfying correct DM relic and direct-detection rate are shown in the plane of $\Delta{M}-M_\text{DM}$. The black colored points are ruled out by $N_{\rm{eff}}$ constraints from Planck 2018 at $2\sigma$ C.L., the maroon points are ruled out by prompt decay bounds from ATLAS, and the cyan-shaded region does not allow inverse first-order EWPT. Benchmark points from Table~\ref{tab:newBPs} are shown using blue stars.}
    \label{fig:result}
\end{figure}

We have studied one of the popular DM scenarios known as the singlet-doublet fermion DM in a framework which can explain light Dirac neutrino mass at one-loop while also affecting the dynamics of electroweak phase transition non-trivially. Two pairs of singlet-doublet fermions, one singlet scalar and three right-handed neutrinos are included as new fields beyond the standard model to generate radiative Dirac neutrino masses. While the lightest neutral mass eigenstate among the SD fermions plays the role of DM, the heavier generation of SD fermion assist in generating an inverse first-order EWPT. Dirac nature of light neutrinos also bring additional detection prospects in the form of dark radiation or $\Delta N_{\rm eff}$ which can be probed at CMB experiments. We identify the parameter space allowed from constraints related to neutrino mass, dark matter, collider, flavor, $\Delta N_{\rm eff}$ and find their detection prospects at ongoing and future experiments. The final allowed parameter space is shown by red colored points in Fig.~\ref{fig:result} in the plane of $\Delta M$ and $M_{\rm DM}$. In the cyan region where \( M_{\chi'} > 2215 \) GeV and the Yukawa coupling remains within perturbative limits, the inverse phase transition scenario does not occur as the corresponding SD fermions are too heavy to affect the electroweak phase transition. The black colored points are ruled out from CMB constraints on $\Delta N_{\rm eff}$ and the maroon colored points are ruled out by prompt decay bounds from ATLAS, whereas the gray shaded region is disfavored from LEP data. The benchmark points given in Table \ref{tab:newBPs}, \ref{tab:newBPsgw} are highlighted as $\star$-shaped points. Interestingly, the heavier SD fermion not only ensure an inverse first-order EWPT, but also leads to another FOPT subsequently when the Universe finally settles to the broken electroweak minima. This leads to two different GW spectra related to two different FOPT with distinct peak frequencies, as shown in Fig.~\ref{fig:gw}. While we have considered the possibility of two such FOPTs for the chosen benchmark points, it is possible to have some regions of parameter space where only one of these transitions is of first order. For such FOPT driven by fermions, one generically requires large fermion-Higgs coupling $\sim \mathcal{O}(1)$. While these couplings are within perturbative limits at the scale of the phase transition, they become non-perturbative at a few TeV scale due to renormalization group evolution (RGE), as shown in Appendix \ref{sec:rge}. This indicates the possibility of new UV completions or strong dynamics at the scale of a few TeV. We leave such possibilities to future studies.

\section*{Acknowledgments}
The authors would like to acknowledge the hospitality at IIT Hyderabad during the WHEPP 2025, where this work was discussed. The work of D.B. is supported by the Science and Engineering Research Board (SERB), Government of India grant CRG/2022/000603. 
\appendix

\section{RGE of couplings}
\label{sec:rge}
In this Appendix, we briefly discuss the RGE running of the couplings in our model. As discussed earlier, the fermion-driven FOPT scenario requires large couplings of singlet-doublet fermions to the Higgs. Even if these couplings remain perturbative at low energy scale, they may become non-perturbative and hit the Landau pole at a higher scale due to RGE running.

We use the package \texttt{SARAH} \cite{Staub:2008uz,Staub:2015kfa} to obtain the one-loop RGE equations for the relevant couplings. In our analysis, we include the running of the singlet-doublet Yukawa coupling, the top Yukawa coupling, and the gauge couplings, while neglecting the Yukawa couplings of the lighter SM fermions since their effects are negligible. The resulting RGE equations are given by
\begin{eqnarray}
    \label{eq:RGE}
    \frac{dy_t}{d(\rm{ln}\mu)}&=&\frac{y_t}{16\pi^2}\left( \frac{9}{2}|y_t|^2+2|y_{\chi}|^2-8 g_3^2-\frac{17}{20}g_1^2-\frac{9}{4}g_2^2\right),\nonumber\\
    \frac{dy_\chi}{d(\rm{ln}\mu)}&=&\frac{y_\chi}{16\pi^2}\left( \frac{7}{2}|y_{\chi}|^2+3|y_t|^2-\frac{9}{20}g_1^2-\frac{9}{4}g_2^2\right),\nonumber\\
    \frac{dg_1}{d(\rm{ln}\mu)}&=&\frac{1}{16\pi^2}\frac{9}{2}g_1^3,\nonumber\\
    \frac{dg_2}{d(\rm{ln}\mu)}&=&\frac{-1}{16\pi^2}\frac{5}{2}g_2^3,\nonumber\\
    \frac{dg_3}{d(\rm{ln}\mu)}&=&\frac{-1}{16\pi^2}7g_3^3.
\end{eqnarray}
\begin{figure}
    \centering
    \includegraphics[scale=0.45]{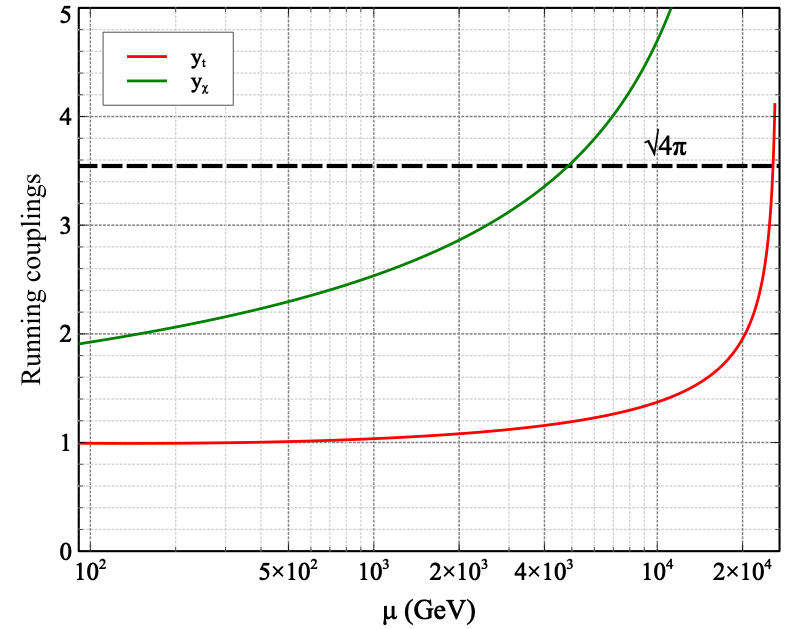}
    \caption{The running top Yukawa and the singlet-doublet Yukawa couplings with energy scale $\mu$ for BP1.}
    \label{fig:RGE01}
\end{figure}
Fig.~\ref{fig:RGE01} shows the running of the top Yukawa coupling and the singlet-doublet Yukawa coupling obtained by solving the RGEs in Eq.~\eqref{eq:RGE}. The boundary conditions are taken as $y_{\chi}(M_{\psi'}) = 2.76$, $y_t(M_{\psi'}) = 1.066$, $g_1(M_{\psi'}) = 0.3608$, $g_2(M_{\psi'}) = 0.6396$, and $g_3(M_{\psi'}) = 1.0357$. As shown in Fig.~\ref{fig:RGE01}, the singlet-doublet Yukawa coupling ($y_\chi$) exceeds the perturbative limit at $\mu \sim 4.85$ TeV and eventually develops a Landau pole at $\mu \sim 26$ TeV. This is a generic issue of such scenarios with fermion driven FOPT and was also noted in earlier works like \cite{Angelescu:2018dkk, Ai:2025vfi}.

\section{Decay rates}\label{app:decayrate}

The calculation of kinematically allowed two and three body decay modes of the charged and neutral component of the doublet are given below.
The two-body decay rates are given as
	\begin{equation}
		\Gamma_{a\rightarrow bc}=\frac{\sqrt{\Lambda(M_a^2,M_b^2,M_c^2)}}{2 M_a}\frac{1}{32 \pi^2M_a^2}4\pi\left|\mathcal{M}_{a\rightarrow bc}\right|^2,
	\end{equation}
    where
	\begin{eqnarray}
        \Lambda(x,y,z)&=&x^2+y^2+z^2-2xy-2yz-2xz, \nonumber \\
		\left|\mathcal{M}_{\psi^0\rightarrow\chi H}\right|^2&=&\frac{(M_{\chi_0}-M_{\chi_1})^2\sin^22\theta}{4v^2}\left(1-2\sin^2\theta\right)^2\left(M_{\chi_1}^2+2M_{\chi_0}M_{\chi_1}+M_{\chi_0}^2-M_h^2\right), \nonumber \\
		\left|\mathcal{M}_{\psi^0\rightarrow\chi Z}\right|^2&=&G_F\left(1-\sin^2\theta\right)\sin^2\theta(M^4_{\chi_1}-2M^2_{\chi_0}M^2_{\chi_1}+M^2_{\chi_1}M^2_Z-6M_{\chi_0}M_{\chi_1}M^2_Z+M^4_{\chi_0}\nonumber\\&&+M^2_{\chi_0}M^2_Z-2M^4_Z), \nonumber \\
		\left|\mathcal{M}_{\psi^-\rightarrow\chi W}\right|^2&=&\frac{G_F}{2}\sin^2\theta(M^4_{\chi_1}-2M^2_{\psi^-}M^2_{\chi_1}+M^2_{\chi_1}M^2_W-6M_{\psi^-}M_{\chi_1}M^2_W+M^4_{\psi^-}\nonumber\\&&+M^2_{\psi^-}M^2_Z-2M^4_Z).
	\end{eqnarray}
	The three-body decay rate is given by \cite{Cirelli:2005uq}
	\begin{eqnarray}
		\Gamma_{\psi\rightarrow\chi l\nu_l}=\frac{2G_F^2}{15\pi^3}(\Delta M )^5 \sin^2\theta.
		\label{eq:3bodydecay}
	\end{eqnarray}

\providecommand{\href}[2]{#2}\begingroup\raggedright\endgroup

\end{document}